\documentclass[twoside,english,reprint]{revtex4-1}
\usepackage{lmodern}
\usepackage[latin9]{inputenc}
\usepackage{babel}
\usepackage{amsmath}
\usepackage{amssymb}
\usepackage{graphicx}
\usepackage{xcolor}
\usepackage{xr-hyper}
\usepackage[colorlinks=true,allcolors=blue]
 {hyperref}

\makeatletter

\usepackage{bbm}

\makeatother

\begin{document}

\title{Remote control of chemistry in optical cavities }

\author{Matthew Du, Raphael F. Ribeiro, Joel Yuen-Zhou$^{*}$}

\address{Department of Chemistry and Biochemistry, University of California
San Diego, La Jolla, California 92093, United States}
\email{joelyuen@ucsd.edu}

\date{\today}
\begin{abstract}
Manipulation of chemical reactivity often involves changing reagents
or environmental conditions. Alternatively, strong coupling between
light and matter offers a way to tunably hybridize their physicochemical
properties and thereby change reaction dynamics without synthetic
modifications to the starting material. Here, we theoretically design
a polaritonic (hybrid photonic-molecular) device that supports ultrafast
tuning of reaction yields even when the catalyst and its reactant
are spatially separated across several optical wavelengths. We demonstrate
how photoexcitation of a `remote catalyst' in an optical microcavity
can control photochemistry of a reactant in another microcavity. Harnessing
delocalization across the spatially separated compounds that arises
from strong cavity-molecule coupling, this intriguing phenomenon is
shown for the infrared-induced \textit{cis} $\rightarrow$ \textit{trans}
conformational isomerization of nitrous acid (HONO). Indeed, increasing
the excited-state population of the remote catalyst can enhance the
isomerization efficiency by an order of magnitude. The theoretical proposal reported herein is generalizable to other reactions and thus introduces a
versatile tool to control photochemistry.   

\end{abstract}
\maketitle
In photochemistry, energy transfer from light to matter produces nonequilibrium
distributions of molecular states, therefore enabling selective initiation
of reactive trajectories. For a given reaction,
tuning of yields is commonly achieved by surveying a series of chemical
analogs. These compounds undergo the same process but on different
potential energy surfaces. The ability to synthesize substrates with
sufficiently varying energetics, though, limits the range of accessible
yields.

\begin{figure}[h]
\includegraphics[scale=0.5]{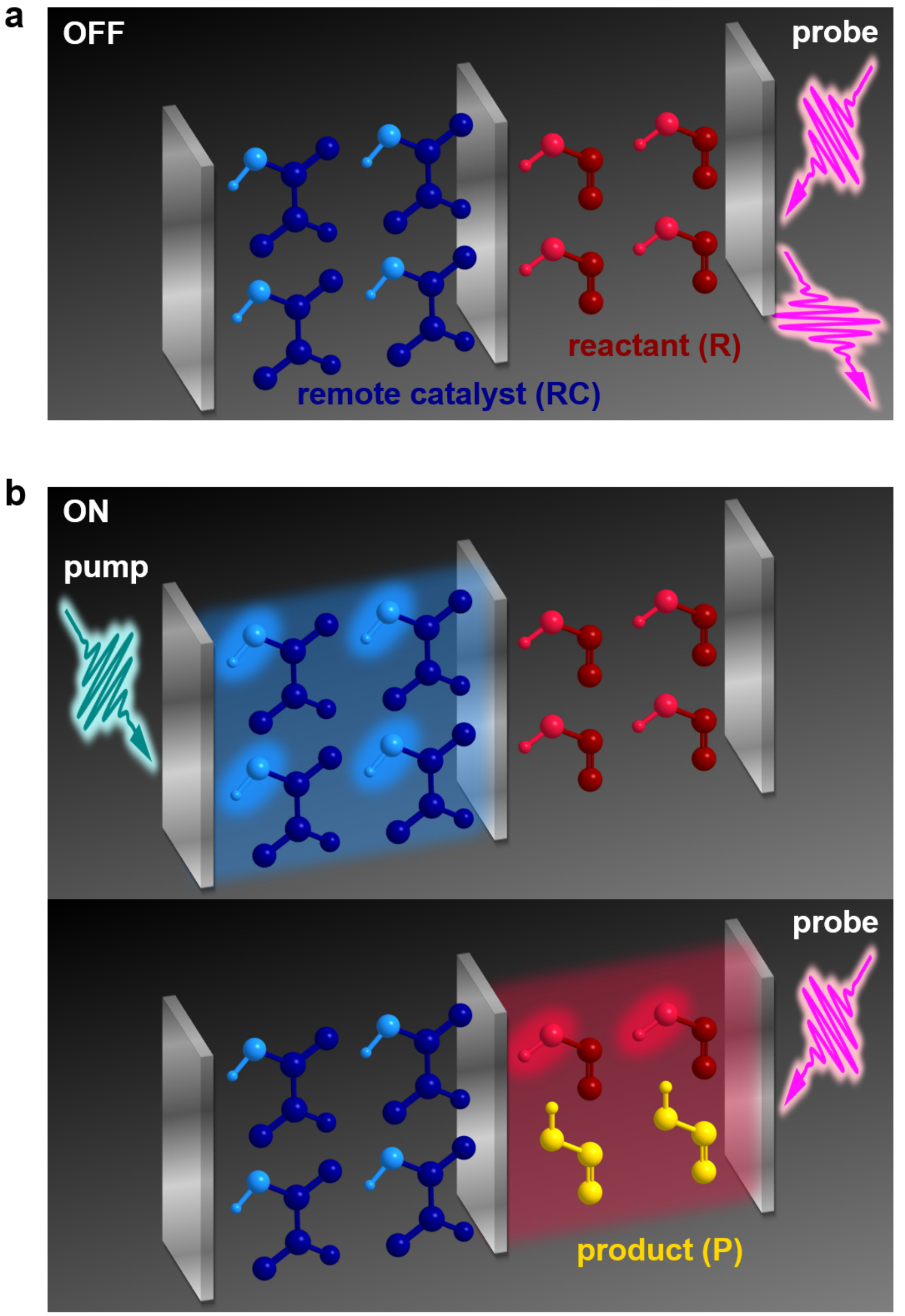}\caption{\textbf{Optical microcavities enable remote control of infrared-induced
conformational isomerization of HONO.} \textbf{a}, Reaction is off.
Without excitation of `remote catalyst' (RC, blue molecule) Tc-glyoxylic
acid, a `probe' laser pulse impinging on the cavity containing reactant
(R, red molecule) \textit{cis}-HONO is off-resonant with polaritons
and thereby reflected; no reaction occurs. \textbf{b}, Reaction is
on. First, a `pump' laser pulse impinging on the mirror of the RC
cavity excites a polariton whose character is predominantly the RC
cavity and the strongly coupled OH stretch (light blue bond) of RC.
Next, the probe pulse is now able to efficiently excite a polariton
whose character is predominantly the R cavity (light red) and the
strongly coupled OH stretch (light red bond) of R; R subsequently
converts into the product molecule (P, yellow molecule) \textit{trans}-HONO.\label{fig:rc_scheme}}
\end{figure}

More facile chemical control of photoinduced reactivity is attainable
in the \textit{strong coupling}\cite{Torma2015} limit. In this regime,
energy coherently oscillates between light and matter faster than
the rates at which their respective excitations decay, and the photonic
and molecular states hybridize into \textit{polariton states}\cite{Agranovich2011}.
To reach sufficiently strong interaction between light and matter,
ensembles of molecules can be placed in optical microcavities (Fig.
\ref{fig:rc_scheme})\cite{Agranovich2011}. These nano- or microstructures
support electromagnetic modes that form polaritons with molecular
superpositions of the same symmetry as the spatial profile of the
electric field. Importantly, the majority of linear combinations of
matter energy states do not possess the right symmetry to mix with
light (in realistic systems, slight mixing occurs due to symmetry-breaking
environments; for example, see ref. \cite{Neumann2018arxiv}) and
constitute the reservoir of \textit{dark states}, which remains centered
at the original molecular transition energy and plays a crucial role
in the relaxation dynamics of polaritons\cite{Agranovich2011,Ebbesen2016,Ribeiro2018rev}.
The energetic consequences and resulting reactivity of molecular polaritons
have seen a surge in interest over the past several years\cite{Ebbesen2016,Feist2017,Ribeiro2018rev,Flick2018review,Schafer2018}.
Since the observation of suppressed conversion between spiropyran
and merocyanine organic dyes\cite{Hutchison2012}, modified kinetics
upon polariton formation have been demonstrated in a wide variety
of photochemical processes by experimental (reverse intersystem crossing\cite{Stranius2018},
photobleaching\cite{Munkhbat2018}, triplet-triplet annihilation\cite{Polak2018arxiv},
water splitting\cite{Shi2018}), theoretical (charge transfer\cite{Herrera2016,Groenhof2018},
dissociation\cite{Kowalewski2016,Vendrell2018}, isomerization\cite{Galego2016},
singlet fission\cite{Martinez-Martinez2018sf}), or both types of
studies (energy transfer\cite{Zhong2017,Du2018,Saez-Blazquez2018,Reitz2018,Schachenmayer2015}).
In addition to detuning the cavity from molecular resonances, polaritonic
systems offer 
a robust control knob of reaction energetics: reactant concentration $N/V$ (more precisely, $N$ is the total number
of cavity-coupled reactant transitions and $V$ is the cavity mode
volume)
\cite{Ebbesen2016,Feist2017,Ribeiro2018rev,Flick2018review}. Indeed, the dependence
of light-matter coupling strength, and the concomitant polaritonic
energy splittings (Fig. \ref{fig:rc_MF}), on $\sqrt{N/V}$ has enabled
concentration-controlled tuning of a number of the aforementioned
processes\cite{Hutchison2012,Zhong2017,Stranius2018}. While robust
compared to substituting the reactant species, changing the concentration
is still prone to issues of unfavorable intermolecular interactions,
particularly insolubility.

Another convenient way to modulate the light-matter coupling is laser-driven
ultrafast population of the dark-state reservoir\cite{Dunkelberger2016,Dunkelberger2018,Xiang2018,Ribeiro2018vp}.
In pump-probe spectroscopy of vibrational polaritons, the pump excitation
of the polaritons is followed by subsequent relaxation into the dark-state
reservoir within $\sim$10-100 ps delay time. This excited-state reservoir,
owing to its large density of almost purely vibrational states, acts
as a very efficient energy sink for the polaritons\cite{delPino2015}.
Due to vibrational anharmonicity, the 1 $\rightarrow$ 2 transitions
are detuned from the 0 $\rightarrow$ 1 transitions and therefore
do not couple as well to cavity modes that are resonant with the latter.
In other words, the concentration $N/V$ of molecular transitions
that can strongly couple to the cavity mode is effectively reduced
on an ultrafast timescale. The reduction is tuned by varying the intensity
of the pump and detected in the frequency-resolved transient transmission
of the probe pulse\cite{Dunkelberger2018,Xiang2018}. 

Here we theoretically demonstrate ultrafast and remote tuning of reaction
yields of the infrared-induced \textit{cis} $\rightarrow$\textit{
trans} fast isomerization channel of nitrous acid (HONO)\cite{Schanz2005}.
Observed in solid Kr matrices using ultrafast spectroscopy, this reaction
is initiated by excitation of the OH stretch vibration of the \textit{cis}
(reactant) conformer. Product formation happens on a 20 ps timescale
with an appreciable quantum yield of 10\%. Therefore, this isomerization
should serve as an ideal candidate to study photoinduced processes
involving vibrational polaritons, given that typical infrared-optical
microcavities are sufficiently long-lived ($\sim$1-10 ps)\cite{Dunkelberger2016,Dunkelberger2018,Xiang2018}
to accommodate the described chemical transformation. We propose a
polaritonic device (Fig. \ref{fig:rc_scheme}) that consists of two
microcavities containing respectively `remote catalyst' (RC) Tc-glyoxylic
acid\cite{Olbert-Majkut2014cpl,Olbert-Majkut2014jpca} (Supplementary
Note \ref{subsec:glyoxylic_acid}) and reactant (R) \textit{cis}-HONO.
Strong coupling exists between the OH stretch ensembles of the molecules
and their corresponding host microcavities, as well as between the
microcavities\cite{Stanley1994}. The resulting polaritonic eigenstates
are delocalized among both RC and R molecules. It follows that without
any direct interaction between the two molecular species, pump-driven
population of the RC dark-state reservoir can modify the energetics
of R and thereby its reactivity in probe-driven conversion to product
(P) \textit{trans}-HONO (Fig. \ref{fig:rc_scheme}, cf. a and b).
Specifically, the probe pulse\textemdash which impinges on the R cavity\textemdash can
be set off-resonant with polaritons with R character such that pumping\textemdash with
a pulse impinging on the RC cavity\textemdash shifts them into resonance
(Fig. \ref{fig:rc_MF}, cf. a and b). By increasing the pump field
intensity, this nonlocal strategy can tune reaction efficiency by
an order of magnitude. 

\subsection*{Results}

To model ultrafast tuning of the photoinitiated R $\rightarrow$ P
conversion, we first describe the bare reaction (\textit{i.e.}, that without
strong light-matter coupling) as comprising three steps. The first
is absorption of light to create a single OH stretch excitation in
R, which we label with $|\text{R}\rangle$ (see Methods). The second
step is intramolecular vibrational redistribution (IVR, \cite{Nesbitt1996})
transition from $|\text{R}\rangle$ to the near-resonant seventh overtone
mode of the torsional coordinate. Given the proximity of this highly
excited state to the barrier of the torsional double-well potential
energy surface and its consequent delocalization across R and P\cite{Schanz2005},
the third step is relaxation into the R and P local well via interaction
with matrix degrees of freedom. For simplicity of notation, we hereby
refer to the product-yielding overtone state as $|\text{P}\rangle$,
although it should be clear that it has mixed character of R and P.
This mechanism is in line with that first proposed for the reaction
induced by pulsed\cite{Schanz2005} and continuous-wave\cite{Hall1963}
excitation, and is in qualitative agreement with mechanisms suggested
by later studies (Supplementary Note \ref{subsec:isomerization}). 

Having addressed the main features of the reaction in the conventional
photochemical setting, we next proceed to describe it within the context
of the proposed device, where the probe absorption into the polariton
states triggers IVR onto $|\text{P}\rangle$. Both absorption and
IVR are treated with a version of input-output theory\cite{Gardiner1985,Ciuti2006,Li2018}
adapted to pump-probe spectroscopy for vibrational polaritons\cite{Ribeiro2018vp}
(Supplementary Note \ref{subsec:reactionSC}). In this approach, the
pump-induced population of dark reservoir states, denoted by an effective
fraction $f_{\text{pump}}$ of the total number of molecules $N$
in the molecular ensemble, controls the nonlinear spectral features.
The major qualitative and quantitative features of experimental transient
spectra are captured within this theory\cite{Xiang2018,Ribeiro2018vp}\textemdash including
the frequencies and intensities of the resonances exhibited by the
transient transmission of the probe. In this work, we disregard electrical
anharmonicity\cite{Ribeiro2018vp} (and fine-structure contributions
such as molecular rotations\cite{Cwik2016,Szidarovszky2018}), whose
inclusion should not qualitatively change our main findings.

\begin{figure*}
\includegraphics[scale=0.75]{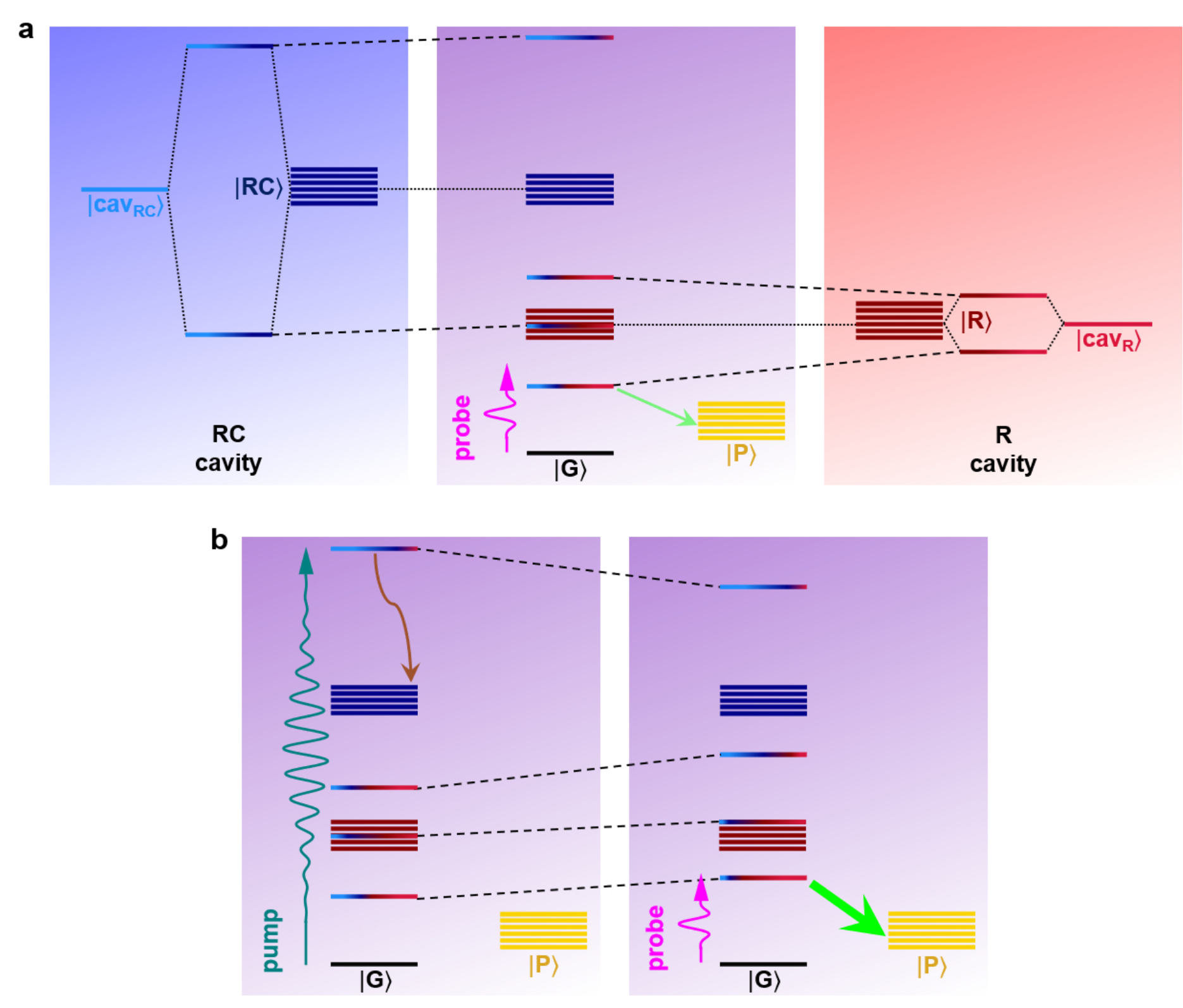}

\caption{\textbf{Pumping of the `remote catalyst' (RC) modifies reaction energetics.}
\textbf{a}, (Blue and red panels) Represented by dotted lines, strong
coupling between cavity photon $|\text{cav}_{\text{RC}}\rangle$ ($|\text{cav}_{\text{R}}\rangle$)
and the OH stretch $|\text{RC}\rangle$ ($|\text{R}\rangle$) in $N$
molecules of RC (reactant, R) produces two polaritons and $N-1$ dark
states; $N=6$ is depicted here for illustration. (Purple panel) Intercavity
coupling further hybridizes (dashed lines) the photonic and vibrational
states into polaritons of the entire light-matter device. A probe
excitation is off-resonant with the lowest polariton, affording inefficient
transfer (thin green arrow) into product-yielding state $|\text{P}\rangle$.
\textbf{b}, (Left panel) Pump excitation of the highest polariton
is followed by relaxation (brown arrow) into the RC dark states during
a \textless{} 100 ps delay time. Owing to anharmonicities, this pumping
reduces the collective $|\text{RC}\rangle$-$|\text{cav}_{\text{RC}}\rangle$
interaction and modifies (dashed lines) the polaritons within an ultrafast
timescale. (Right panel) The probe excitation is now resonant with
the lowest polariton, which has increased mixing fraction of R and
its cavity, affording efficient transfer (thick green arrow) into
$|\text{P}\rangle$. In both \textbf{a} and \textbf{b}, $|\text{G}\rangle$
is the molecular and photonic vacuum ground state (energy not drawn
to scale), and the polariton vertical positions (color gradients)
represents their energies (mixing fractions) (see Supplementary Note
\ref{subsec:MF}).\label{fig:rc_MF}}
\end{figure*}

Before proceeding to analyze the dynamics of the remote-control device,
we investigate its spectral features. In the first excitation manifold,
the Hamiltonian for the polariton states in the basis of the constituent
species is ($\hbar=1$)
\begin{equation}
H_{\text{system}}^{(\text{no pump})}=\begin{pmatrix}\omega_{\text{RC}} & g_{\text{RC}}\sqrt{N_{\text{RC}}} & 0 & 0\\
g_{\text{RC}}\sqrt{N_{\text{RC}}} & \omega_{\text{cav}_{\text{RC}}} & g_{\text{cav}} & 0\\
0 & g_{\text{cav}} & \omega_{\text{cav}_{\text{R}}} & g_{\text{R}}\sqrt{N_{\text{R}}}\\
0 & 0 & g_{\text{R}}\sqrt{N_{\text{R}}} & \omega_{\text{R}}
\end{pmatrix},\label{eq:h0hybrid}
\end{equation}
where the entries from left to right (top to bottom) represent the
OH stretch excitation in RC (labeled by $|\text{RC}\rangle$), the
cavity hosting RC, the cavity hosting R, and $|\text{R}\rangle$,
respectively. For simplicity, we take each cavity mode to be resonant
with the hosted OH vibration: $\omega_{\text{cav}_{\text{RC}}}=\omega_{\text{RC}}=3455\text{ cm}^{-1}$
(the average frequency of the OH stretch of RC\cite{Olbert-Majkut2014jpca,Olbert-Majkut2014cpl},
see Supplementary Note \ref{subsec:glyoxylic_acid}) and $\omega_{\text{cav}_{\text{R}}}=\omega_{\text{R}}=3402\text{ cm}^{-1}$
(the frequency of the OH stretch vibration of R\cite{Schanz2005}).
Let the collective light-matter couplings\cite{Ribeiro2018rev} be
$g_{\text{RC}}\sqrt{N_{\text{RC}}}=57\text{ cm}^{-1}$ and $g_{\text{R}}\sqrt{N_{\text{R}}}=11\text{ cm}^{-1}$
for numbers $N_{\text{RC}}$ and $N_{\text{R}}$ of RC and R molecules,
respectively, per mode volume $V$ of each cavity; we have incorporated
$V$ into the $g$ values for notational convenience. These couplings
correspond to $\sim1.7\%$ and $\sim0.3\%$ of the transition energy
of each interacting species, comparable to experimental values (0.2-2.2\%)
for mid-infrared vibrational polaritons\cite{Dunkelberger2018,Casey2016}.
With no cavity-cavity coupling ($g_{\text{cav}}=0$; Fig. \ref{fig:rc_MF}a,
blue and red panels), one pair of polaritons has character of only
RC and its cavity (blue), and the other pair only R and its cavity
(red). Upon introduction of intercavity coupling $g_{\text{cav}}=27\text{ cm}^{-1}$
(Fig. \ref{fig:rc_MF}a, purple panel)\textemdash corresponding to
$\sim0.8\%$ of the cavity photon energies\cite{Stanley1994}\textemdash the
three lowest polaritons from the $g_{\text{cav}}=0$ case hybridize
into three states delocalized across RC, R, and their host cavities.
This delocalization enables pumping of RC to remotely tune the energy
of polaritons with R character and thereby the isomerization efficiency.
In contrast, the highest polariton for $g_{\text{cav}}=0$ is spectrally
isolated and does not change much in energy or character when intercavity
coupling is introduced (Fig. \ref{fig:rc_MF}a, cf. blue and purple
panels).

Nevertheless, population of the dark RC states is achievable via excitation
of this highest polariton with a pump pulse (Fig. \ref{fig:rc_MF}b,
left panel) impinging on the RC cavity (Fig. \ref{fig:rc_scheme}b,
top panel). Given large enough $g_{\text{RC}}$ and $\omega_{\text{RC}}-\omega_{\text{R}}$,
this highest level is essentially half RC and half RC cavity in character.
Furthermore, by conservation of the number of energy levels, there
are $N_{\text{RC}}-1$ RC dark reservoir states, significantly larger
than 4, the number of polaritons. Downhill energy relaxation from
the highest polariton is then most likely to occur into the relatively
dense RC dark manifold\cite{Agranovich2011,delPino2015,Ribeiro2018rev}.
In fact, this process is permitted in a matrix of Kr (Supplementary
Fig. \ref{fig:SI_rc_pump}). 

The resulting pump-dependent effective Hamiltonian is\cite{Ribeiro2018vp}\begin{widetext}

\begin{equation}
H_{\text{system}}^{(\text{pump})}=\begin{pmatrix}\omega_{\text{RC}} & 0 & g_{\text{RC}}\sqrt{(1-2f_{\text{pump}})N_{\text{RC}}} & 0 & 0\\
0 & \omega_{\text{RC}}+2\Delta & g_{\text{RC}}\sqrt{2f_{\text{pump}}N_{\text{RC}}} & 0 & 0\\
g_{\text{RC}}\sqrt{(1-2f_{\text{pump}})N_{\text{RC}}} & g_{\text{RC}}\sqrt{2f_{\text{pump}}N_{\text{RC}}} & \omega_{\text{cav}_{\text{RC}}} & g_{\text{cav}} & 0\\
0 & 0 & g_{\text{cav}} & \omega_{\text{cav}_{\text{R}}} & g_{\text{R}}\sqrt{N_{\text{R}}}\\
0 & 0 & 0 & g_{\text{R}}\sqrt{N_{\text{R}}} & \omega_{\text{R}}
\end{pmatrix}.\label{eq:hpumphybrid}
\end{equation}
\end{widetext} (see Supplementary Note \ref{subsec:d_eff} for derivation
and interpretation). Though not utilized in calculations of absorption
or reaction efficiency, this matrix provides physical intuition in
that it characterizes the polaritonic transitions in the absence of
lineshape broadening (our calculations \textit{do} account for dissipative
effects, see Methods): $|\text{RC}\rangle$, the $|\text{RC}\rangle$
1 $\rightarrow$ 2 transition, the cavity with RC, the cavity with
R, and $|\text{R}\rangle$, from left to right (top to bottom). $\Delta=-89\text{ cm}^{-1}$
is the mechanical anharmonicity of the OH stretch of RC\cite{Olbert-Majkut2014cpl,Olbert-Majkut2014jpca}
(Supplementary Note \ref{subsec:glyoxylic_acid}). It is evident from
equation (\ref{eq:hpumphybrid}) that as the parameter $f_{\text{pump}}$
representing the degree of population of RC dark reservoir states
is increased, the coupling between RC and the cavity is reduced, formalizing
the qualitative arguments provided above. Taking the perspective that
the hybrid states for $g_{\text{cav}}\neq0$ are formed by mixing
the polariton states of each cavity (when $g_{\text{cav}}=0$, Fig.
\ref{fig:rc_MF}a), pumping blueshifts the lower RC polariton away
from the lower R polariton and reduces their mixing when $g_{\text{cav}}\neq0$
(Fig. \ref{fig:rc_MF}b, cf. left and right panels). Indeed, the lowest
polariton for $g_{\text{cav}}\neq0$ becomes predominantly R and its
corresponding cavity upon pumping (Fig. \ref{fig:rc_scheme}b, right
panel). 

\begin{figure}[th]
\includegraphics[scale=0.3]{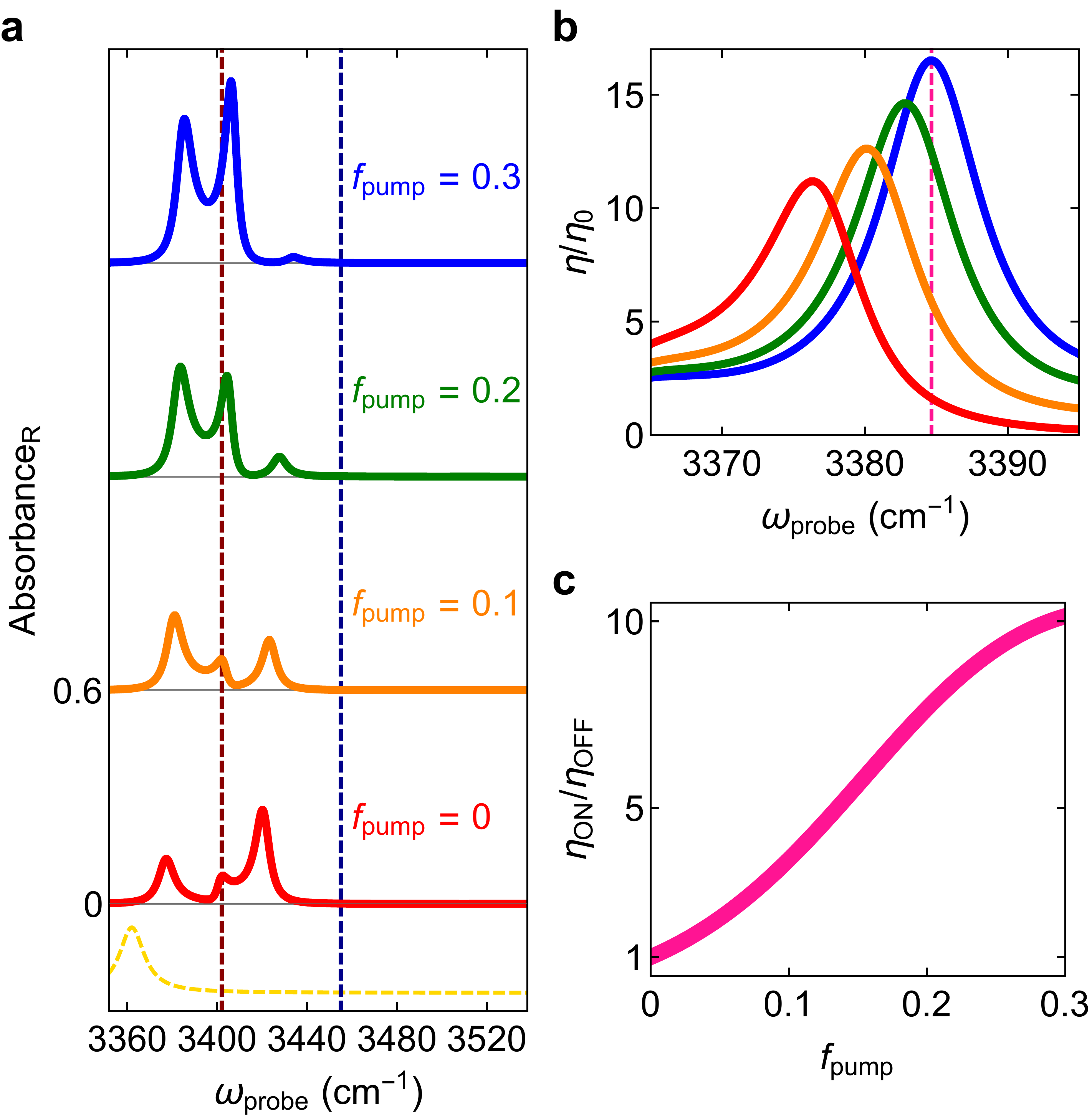}

\caption{\textbf{Pumping of the `remote catalyst' (RC) turns on reaction with
high tunability.} \textbf{a}, Probe energy absorption into reactant
(R) given various fractions $f_{\text{pump}}$ of pump-excited RC
molecules. The dashed lines indicate the energies of the bare vibrations
(\textit{i.e.}, those without strong light-matter coupling) of RC
(dark blue) and R (dark red), as well as the arbitrarily scaled effective
energy lineshape of the non-absorbing quantum state $|\text{P}\rangle$
(yellow) that first receives population from the strongly coupled
R vibration via IVR and then relaxes into product (P) states (see
main text). \textbf{b}, Relative reaction efficiency $\eta/\eta_{0}$
(\textit{i.e.}, compared to the bare efficiency) as a function of
the `probe' pulse energy. The color scheme for the solid lines follows
that of \textbf{a}. \textbf{c}, Enhancement $\eta_{\text{ON}}/\eta_{\text{OFF}}$
of reaction efficiency (\textit{i.e.}, compared to $f_{\text{pump}}=0$)
as a function of $f_{\text{pump}}$ at the probe energy ($\omega_{\text{probe}}\approx3385\text{ cm}^{-1}$)
is indicated by the pink dashed line in \textbf{c}.\label{fig:rc_ON}}
\end{figure}

The ability to shift its energy and increase its R character with
pumping of RC renders this lowest polariton state promising for pump-enhancing
the energy that eventually dissipates into R molecules, and thus the
isomerization triggered when a probe pulse impinges on the R cavity
(Fig. \ref{fig:rc_MF}, cf. a and b). Spectra (Fig. \ref{fig:rc_ON}a)
computed from input-output theory (see Methods) reveal that probe
absorption of the lowest polariton into R is stronger with more pumping.
This trend is in agreement with the pump dependence of its R and R-cavity
mixing fractions (Fig. \ref{fig:rc_MF}b, cf. left and right panels).
As a consistency check, we note that the energies and intensities
of the other absorption peaks (Fig. \ref{fig:rc_ON}a) also agree
with the energies and R/R-cavity characters, respectively (Fig. \ref{fig:rc_MF}).

Now we show that the reaction efficiency $\eta$ is highly pump-tunable
(Fig. \ref{fig:rc_ON}b). Also calculated from input-output theory
(see Methods), $\eta$ (equation (\ref{eq:eta})) is the product of
the probe absorbance into R and the quantum yield of isomerization
from the excited polariton. Given that the peak absorption of the
lowest polariton moves away in energy from $|\text{P}\rangle$ ($\omega_{\text{P}}=3362\text{ cm}^{-1}$\cite{Schanz2005},
Supplementary Note \ref{subsec:isomerization}) with pumping (Fig.
\ref{fig:rc_ON}a), it is somewhat counterintuitive that the corresponding
peak value of $\eta$ increases. This behavior arises because for
all values of $f_{\text{pump}}$, spectral overlap between any polariton
and $|\text{P}\rangle$ is small (Fig. \ref{fig:rc_ON}a). Thus, the
peak $\eta$ is controlled by the position and intensity of the lowest-polariton
absorption. It therefore makes sense that the maximum $\eta$ blueshifts
and rises with higher pumping (Fig. \ref{fig:rc_ON}b). Relative to
the experimental bare reaction efficiency $\eta_{0}$ (see Methods),
the $\eta$ values grow to an order of magnitude greater with increasing
$f_{\text{pump}}$ (Fig. \ref{fig:rc_ON}b). The reason for such high
values is that the lowest polariton is a more efficient absorber (Fig.
\ref{fig:rc_ON}a) and is nearer in resonance to $|\text{P}\rangle$
compared to the bare $|\text{R}\rangle$ (peak absorbance = 0.07\cite{Schanz2005}).
To realize remote tuning of reactivity, we focus on the probe frequency
(Fig. \ref{fig:rc_ON}b, pink dashed line) that corresponds to the
peak $\eta$ for the highest explored fraction $f_{\text{pump}}=0.3$
of pump-excited RC molecules. Notably, pumping enhances the reaction
efficiency $\eta_{\text{ON}}$ for this choice of $\omega_{\text{probe}}$
by over an order of magnitude compared to the efficiency $\eta_{\text{OFF}}$
with no pumping (Fig. \ref{fig:rc_ON}c). As an aside, while uphill
relaxation may happen from the lowest polariton to the dark R states,
this interfering process can be minimized with lower temperatures.
Even if the relaxation is significantly fast, \textit{e.g.}, compared
to polariton decay, the isomerization and its enhancement can still
be observed as long as polariton absorption is detectable. 

\begin{figure}[th]
\includegraphics[scale=0.3]{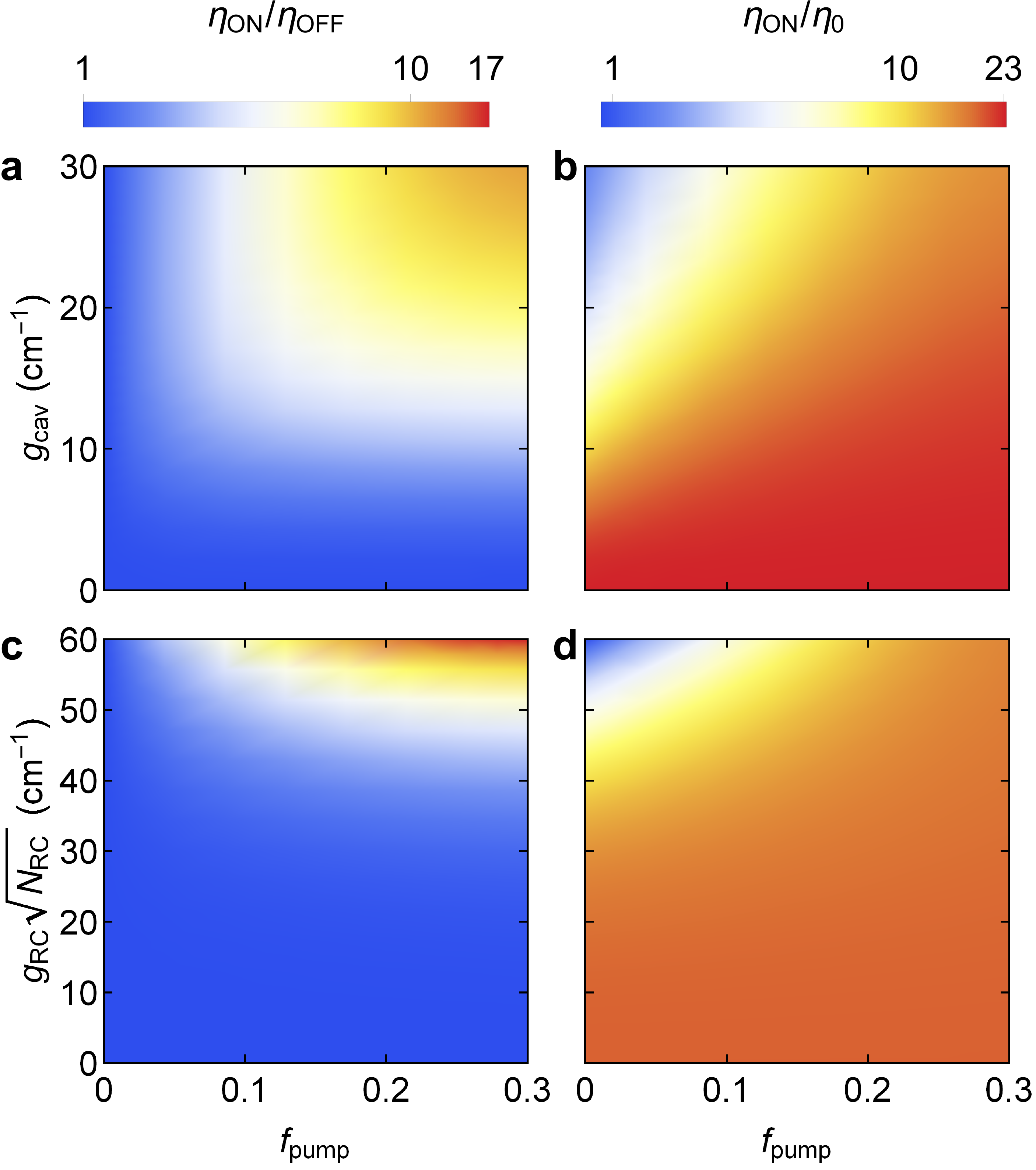}\caption{\textbf{Pump-induced reaction enhancement can be remotely tuned by
varying properties of the `remote catalyst' (RC) and its host cavity.}
Reaction efficiencies $\eta_{\text{ON}}$ relative to those without
pumping ($\eta_{\text{OFF}}$) or in the bare case (\textit{i.e.},
without strong light-matter coupling, $\eta_{0}$), respectively,
as a function of either intercavity coupling strength $g_{\text{cav-cav}}$
(\textbf{a} and \textbf{b}) or collective cavity-RC coupling $g_{\text{RC}}\sqrt{N_{\text{RC}}}$
(\textbf{c} and \textbf{d}) and fraction $f_{\text{pump}}$ of pump-excited
RC molecules. In \textbf{b} and \textbf{d}, notice the ability to
control reaction efficiency even in the linear optical regime ($f_{\text{pump}}=0$).\label{fig:rc_g}}
\end{figure}

So far, we have considered modifying a reaction by optical pumping
of RC. We now briefly show further manipulation of the R $\rightarrow$
P isomerization efficiency via tuning of intercavity and cavity-RC
couplings, both adjustable without \textit{any} direct alteration
of R. Thus, we also show remote control of the reaction in the linear
probe-excitation regime, \textit{i.e.}, without pumping. When the
intercavity coupling is manipulated (Fig. \ref{fig:rc_g}a), \textit{e.g.},
by varying the thickness of the middle mirror\cite{Skolnick1998},
the maximum boost in reaction efficiency with pumping ($\eta_{\text{ON}}$)
over that without ($\eta_{\text{OFF}}$) reaches several tenfold as
$g_{\text{cav}}$ rises from 0 to $30\text{ cm}^{-1}$. Such increase
occurs because delocalization of the lowest polariton across RC, R,
and their host cavities\textemdash and therefore potential for reactive
modification\textemdash{} increases with intercavity coupling. Changing
$g_{\text{cav}}$ raises too the absolute $\eta_{\text{ON}}$ (compared
to bare efficiency $\eta_{0}$, Fig. \ref{fig:rc_g}b) by over an
order of magnitude for fixed $f_{\text{pump}}$, especially when $f_{\text{pump}}=0$.
Alternatively, $g_{\text{RC}}\sqrt{N_{\text{RC}}}$ can be tuned (Fig.
\ref{fig:rc_g}c), \textit{e.g.}, by increasing the concentration
of RC, to yield similar favorable pump-enhancements (Fig. \ref{fig:rc_g},
cf. a and c) and absolute efficiencies (Fig. \ref{fig:rc_g}, cf.
b and d) for fixed pumping. Indeed, cavity-RC coupling too can regulate
the efficiency of the single-pulse photoisomerization (Fig. \ref{fig:rc_g}b,
$f_{\text{pump}}=0$). Notice though that $g_{\text{RC}}\sqrt{N_{\text{RC}}}$
must exceed $\sim40\text{ cm}^{-1}$ to appreciably influence $\eta_{\text{ON}}/\eta_{\text{OFF}}$
(Fig. \ref{fig:rc_g}c) and $\eta_{\text{ON}}/\eta_{0}$ (Fig. \ref{fig:rc_g}d).
The origin of this requirement is the same as that of the pump-induced
modulation (with fixed cavity-RC coupling): adjusting $g_{\text{RC}}\sqrt{N_{\text{RC}}}$
changes the mixing between the polaritons of each cavity (when $g_{\text{cav}}=0$);
control of reactivity is realizable when the lowest polariton of the
entire device is sufficiently delocalized across the photonic and
vibrational species associated with both RC and R. While pumping of
RC provides a very versatile tuning mechanism, in the absence of ultrafast
equipment and so long as increasing the thickness of the intercavity
mirror or the concentration of RC is feasible, the linear optical
experiments suggested provide an interesting alternative to our original
proposal.

\subsection*{Discussion}

We have theoretically demonstrated ultrafast remote control of the
isomerization of \textit{cis}-HONO to \textit{trans}-HONO using an
infrared polaritonic device. The proposed setup consists of two strongly
interacting microcavities containing separate ensembles of `remote
catalyst' (RC) and reactant. The polaritons of the hybrid system are
delocalized across both molecular species and their host cavities.
Acting on the RC cavity, a pump pulse excites the highest polariton,
followed by picosecond-timescale relaxation to the dark-state reservoir
of RC. Due to anharmonicity, the 1 $\rightarrow$ 2 vibrational transitions
of RC are significantly detuned from the RC cavity mode, hence inducing
an effective weakening of the collective coupling (and hybridization)
between RC and the remaining components of the device. The lowest
polariton concomitantly acquires less character of RC and its respective
cavity and more of reactant and its respective cavity. As a result,
probe-pulse excitation acting on the reactant cavity yields enhanced
efficiency of IVR into the product state compared to the no-pumping
case. By raising the pump intensity, the reaction efficiency can be
boosted by an order of magnitude. Remarkably, this tunability requires
no spatial contact whatsoever between RC and reactant, challenging
the paradigm of traditional chemical catalysis. We emphasize that
additional manipulation of reactivity can be achieved by varying the
intercavity or RC-cavity coupling strengths, \textit{e.g.}, by changing
the distance between cavities or the concentration of RC, respectively.
These adjustments extend remote control to the linear optical regime. 

Although our results involve tuning a vibrational excited state that
couples into the reaction coordinate, they can be generalized to electronic
excited states, which feature a variety of photochemical reactions,
some of which have been explored already in the polaritonic regime\cite{Ebbesen2016,Feist2017,Ribeiro2018rev,Flick2018review}.
Success of the proposed strategies relies essentially on (1) the ability
to couple RC and R to interacting cavity modes and (2) a difference
in coupling between the fundamental and anharmonic transition of the
former compound. Indeed, inorganic\cite{Khitrova1999} and organic\cite{Vasa2013,Liu2015}
excitons are satisfactory platforms for realization of the polaritonic
device and the pump-dependent modulation of light-matter coupling
studied here. Furthermore, remote control of reactivity can be extended
to include plasmonic nanostructures, which have been well-studied
in the strong coupling regime\cite{Bellessa2004,Torma2015,Sukharev2017}
and also offer promising routes for ultrafast manipulation of nanoparticle-reactant\cite{Fofang2011},
plasmon-plasmon\cite{Prodan2003}, and photon-plasmon interactions\cite{Thakkar2017}. 

Beyond the application described here, the polaritonic device can
be employed as a diagnostic tool for reaction mechanisms. For example,
identification of states that afford high reactive tunability, when
strongly coupled in the polaritonic device, can provide mechanistic
insight. Such functionality would be especially attractive for processes
that involve a series of state-to-state transitions, \textit{e.g.},
IVR or IVR-driven reactions such as the HONO isomerization studied
here. More broadly, the proposed remote control represents a new class of molecular quantum technologies featuring manipulation of chemical processes through coherent interactions\cite{ShapiroBook}.
In addition, this technique to control reactions without direct catalyst-reactant interaction paves way for novel and possibly greener approaches to catalytic and separations chemistry.

\subsection*{Methods}

\subsubsection*{Hamiltonian for polaritonic device}

The Hamiltonian for the proposed polaritonic setup (Fig. \ref{fig:rc_scheme})
is $H=H_{\text{system}}+H_{\text{bath}}+H_{\text{system}-\text{bath}}$,
where the latter two terms are provided in Supplementary Note \ref{subsec:reactionSC},
and 
\begin{align}
H_{\text{system}} & =H_{\text{RC}}+H_{\text{cav}_{\text{RC}}}+H_{\text{cav}_{\text{RC}}-\text{RC}}\nonumber \\
 & \quad+H_{\text{cav}_{\text{RC}}-\text{cav}_{\text{R}}}+H_{\text{R}}+H_{\text{cav}_{\text{R}}}+H_{\text{cav}_{\text{R}}-\text{R}}\nonumber \\
 & \quad+H_{\text{P}}+H_{\text{R}-\text{P}},\label{eq:hS}
\end{align}
from which Hamiltonian $H_{\text{system}}^{(\text{no pump})}$, equation
(\ref{eq:h0hybrid}), follows in the perturbative limit of $H_{\text{R}-\text{P}}$.
From left to right and top to bottom, the terms of equation (\ref{eq:h0hybrid})
respectively represent the contributions of RC, RC cavity, their coupling,
intercavity coupling, R cavity, R, cavity-$|\text{R}\rangle$ coupling,
P, and the $|\text{R}\rangle$-$|\text{P}\rangle$ IVR coupling. Explicitly,
these terms read ($\hbar=1$)\begin{subequations}\label{eq:hRCall}
\begin{align}
H_{\text{RC}} & =\omega_{\text{RC}}\sum_{i=1}^{N_{\text{RC}}}a_{\text{RC},i}^{\dagger}a_{\text{RC},i}\label{eq:hRC}\\
 & \quad+\Delta\sum_{i=1}^{N_{\text{RC}}}a_{\text{RC},i}^{\dagger}a_{\text{RC},i}^{\dagger}a_{\text{RC},i}a_{\text{RC},i},\label{eq:hRCanh}
\end{align}
\end{subequations}

\begin{align}
H_{\text{cav}_{\text{RC}}} & =\omega_{\text{cav}_{\text{RC}}}c_{\text{RC}}^{\dagger}c_{\text{RC}},\\
H_{\text{cav}_{\text{RC}}-\text{RC}} & =g_{\text{RC}}\sum_{i=1}^{N_{\text{RC}}}(a_{\text{RC},i}^{\dagger}c_{\text{RC}}+\text{h.c.}),\\
H_{\text{cav}_{\text{RC}}-\text{cav}_{\text{R}}} & =g_{\text{cav}}(c_{\text{R}}^{\dagger}c_{\text{RC}}+\text{h.c.}),\\
H_{\text{R}} & =\omega_{\text{R}}\sum_{i=1}^{N_{\text{R}}}a_{\text{R},i}^{\dagger}a_{\text{R},i},\label{eq:hR}\\
H_{\text{cav}_{\text{R}}} & =\omega_{\text{cav}_{\text{R}}}c_{\text{R}}^{\dagger}c_{\text{R}},\label{eq:hcR}\\
H_{\text{P}} & =\omega_{\text{P}}\sum_{i=1}^{N_{\text{R}}}|\text{\ensuremath{\text{P}_{i}\rangle}}\langle\text{P}_{i}|,\label{eq:hP}\\
H_{\text{R}-\text{P}} & =V_{\text{R}-\text{P}}\sum_{i=1}^{N_{\text{R}}}(|\text{P}_{i}\rangle\langle\text{G}|a_{\text{R},i}+\text{h.c.}).\label{eq:hRP}
\end{align}
Here, $a_{x,i}^{\dagger}$ ($a_{x,i}$) is the bosonic creation (annihilation)
operator for an OH stretch excitation at the $i$th molecule of the
$x=\text{RC},\,\text{R}$ species; $c_{x}^{\dagger}$ ($c_{x}$) is
the bosonic creation (annihilation) operator for a photon in the cavity
hosting $x$. Finally, $|\text{P}_{i}\rangle$ is the seventh overtone
(eighth excited state) of HONO molecule $i$ that has mixed \textit{cis}
and \textit{trans} character, and $|\text{G}\rangle$ is the molecular
and photonic vacuum ground state. All energy and coupling parameters
are defined in the main text, except for $V_{\text{R}-\text{P}}$
(see Supplementary Note \ref{subsec:reaction}). Given that population
of just RC (but not R) is assumed to be excited by the pump, only
its anharmonicity is relevant (see equation (\ref{eq:hRCanh})). 

\subsubsection*{Input-output theory for simulating absorption and reaction efficiency}

Following conventional input-output theory\cite{Gardiner1985,Ciuti2006,Li2018},
as well as its adaptation to pump-probe spectroscopy of vibrational
polaritons\cite{Ribeiro2018vp}, we write the Heisenberg-Langevin
equations of motion for the probe-induced dynamics of the proposed
polariton device (see Supplementary Note \ref{subsec:reactionSC}
for derivation and Supplementary Fig. \ref{fig:SI_io_rc} for schematic
representation of the equations):\begin{subequations}\label{eq:HL}
\begin{align}
\frac{dP_{\text{RC}}(t)}{dt} & =-i(\omega_{\text{RC}}-i\gamma_{\text{RC}}/2)P_{\text{RC}}(t)\nonumber \\
 & \quad-ig_{\text{RC}}\sqrt{N_{\text{RC}}}c_{\text{RC}}(t)-2i\Delta P_{\text{RC},3}(t),\label{eq:pRC1}\\
\frac{dP_{\text{RC},3}(t)}{dt} & =-i(\omega_{\text{RC}}+2\Delta-i3\gamma_{\text{RC}}/2)P_{\text{RC},3}(t)\nonumber \\
 & \quad-2ig_{\text{RC}}f_{\text{pump}}\sqrt{N_{\text{RC}}}c_{\text{RC}}(t),\label{eq:pRC3}\\
\frac{dc_{\text{RC}}(t)}{dt} & =-i(\omega_{\text{cav}_{\text{RC}}}-i\kappa_{\text{RC}}/2)c_{\text{RC}}(t)\nonumber \\
 & \quad-ig_{\text{cav}}c_{\text{R}}(t)-ig_{\text{RC}}\sqrt{N_{\text{RC}}}P_{\text{RC}}(t),\label{eq:cRC}\\
\frac{dc_{\text{R}}(t)}{dt} & =-i(\omega_{\text{cav}_{\text{R}}}-i\kappa_{\text{R}}/2)c_{\text{R}}(t)-ig_{\text{cav}}c_{\text{RC}}(t)\nonumber \\
 & \quad-ig_{\text{R}}\sqrt{N_{\text{R}}}P_{\text{R}}(t)-\sqrt{\kappa_{\text{R}}}c_{\text{R},\text{in}}(t),\label{eq:cR}\\
\frac{dP_{\text{R}}(t)}{dt} & =-i(\omega_{\text{R}}-i\gamma_{\text{R}}/2)P_{\text{R}}(t)\nonumber \\
 & \quad-ig_{\text{R}}\sqrt{N_{\text{R}}}c_{\text{R}}(t)-iV_{\text{R}-\text{P}}P_{\text{P}}(t),\label{eq:pR}\\
\frac{dP_{\text{P}}(t)}{dt} & =-i[\omega_{\text{P}}-i(\Gamma_{cis}+\Gamma_{trans})/2]P_{\text{P}}(t)\nonumber \\
 & \quad-iV_{\text{R}-\text{P}}P_{\text{R}}(t).\label{eq:pP}
\end{align}
\end{subequations}Here, $P_{x}=\sum_{i=1}^{N_{x}}a_{x,i}/\sqrt{N_{x}}$
is the linear molecular polarization representing the collective bright
molecular states $|\text{R}\rangle=P_{\text{R}}^{\dagger}|\text{G}\rangle$
and $|\text{RC}\rangle=P_{\text{RC}}^{\dagger}|\text{G}\rangle$ mentioned
in the main text. $P_{x}$ is coupled to cavity polarization $c_{x}^{\dagger}$;
$P_{\text{RC},3}=\sum_{i=1}^{N_{\text{RC}}}a_{\text{RC},i}^{\dagger}a_{\text{RC},i}a_{\text{RC},i}/\sqrt{N_{\text{RC}}}$
is the third-order polarization for RC that depends on the pump-induced
excited-state fraction $f_{\text{pump}}$, a parameter that describes
the extent of pumped RC population stored in the corresponding dark-state
reservoir (see main text). The decay constants $\gamma_{\text{RC}}=\gamma_{\text{R}}=5\text{ cm}^{-1}$
approximate the absorption linewidth of the OH stretch excitation
in R\cite{Schanz2005} and RC\cite{Olbert-Majkut2014jpca,Olbert-Majkut2014cpl}.
The cavity photon lifetimes $\kappa_{\text{RC}}=\kappa_{\text{R}}=9.5\text{ cm}^{-1}$
are chosen such that $2\kappa_{x}/\omega_{x}$ (each cavity has only
one mirror that couples to external photons, Fig. \ref{fig:rc_scheme})
approximately matches experimental parameters\cite{Ribeiro2018vp,Xiang2018}.
We have also defined the operator $P_{\text{P}}=\sum_{i=1}^{N_{\text{R}}}|\text{G}\rangle\langle\text{P}_{i}|/\sqrt{N_{\text{R}}}$
which keeps track of the IVR transferred population from $|\text{R}\rangle$
to $|\text{P}\rangle=P_{\text{P}}^{\dagger}|\text{G}\rangle$. The
remaining decay rates $\Gamma_{cis}$ and $\Gamma_{trans}$ (Supplementary
Note \ref{subsec:reaction}) represent relaxation of the HONO torsional
state into the local R (\textit{cis}) or P (\textit{trans}) wells.
The operator $c_{\text{R},\text{in}}$ represents the external probe
field which couples into the system via the R cavity. 

Carrying out the Fourier transform of equation (\ref{eq:HL}), treating
IVR as a perturbation in $V_{\text{R}-\text{P}}$ (Supplementary Note
\ref{subsec:reactionSC}) and solving for $S_{\text{R}}(f_{\text{pump}},\omega)$
satisfying $P_{\text{R}}(\omega)=S_{\text{R}}(f_{\text{pump}},\omega)c_{\text{R},\text{in}}(\omega)$
allows for simulation (\textit{e.g.}, Fig. \ref{fig:rc_ON}a in the
main text) of the probe absorbance by R:
\begin{equation}
\text{absorbance}_{\text{R}}(f_{\text{pump}},\omega)=\gamma_{\text{R}}|S_{\text{R}}(f_{\text{pump}},\omega)|^{2}.\label{eq:abs_formula}
\end{equation}
Doing the same steps with the P equation of motion gives the reaction
efficiency
\begin{equation}
\eta(f_{\text{pump}},\omega)=\text{absorbance}_{\text{R}}(f_{\text{pump}},\omega)\text{QY}_{\text{R}\rightarrow\text{P}}(\omega),\label{eq:eta}
\end{equation}
 where (Supplementary Note \ref{subsec:reaction})
\begin{align}
\text{QY}_{\text{R}\rightarrow\text{P}}(\omega) & =\frac{\gamma_{\text{R}\rightarrow\text{P}}(\omega)}{\gamma_{\text{R}}}
\end{align}
 is an isomerization quantum yield, and 
\begin{equation}
\gamma_{\text{R}\rightarrow\text{P}}(\omega)=\frac{V_{\text{R}\rightarrow\text{P}}^{2}\Gamma_{trans}}{(\omega-\omega_{\text{P}})^{2}+[(\Gamma_{cis}+\Gamma_{trans})/2]^{2}}\label{eq:gRPmethods}
\end{equation}
is the transition rate from $|\text{R}\rangle$ to $|\text{P}\rangle$.
The $\eta$ values discussed in the main text are defined as
\begin{align}
\eta_{\text{ON}} & =\eta(0.3,\omega_{\text{ON}}),\\
\eta_{\text{OFF}} & =\eta(0,\omega_{\text{ON}}),\\
\eta_{0} & \equiv\eta_{0}(\omega_{\text{R}}).\label{eq:eta0mt}
\end{align}
$\omega_{\text{ON}}$ (Fig. \ref{fig:rc_ON}b, pink dashed line) is
the frequency that maximizes $\eta(0.3,\omega)$ in the region $\omega\in[\omega_{\text{P}}-2\text{ cm}^{-1},\omega_{\text{R}}-g_{\text{R}}\sqrt{N_{\text{R}}}+2\text{ cm}^{-1}]$,
containing the lowest polariton lineshape for all coupling strengths
explored in this work. The derivation of $\eta_{0}(\omega)$ is given
in Supplementary Note \ref{subsec:reaction}. 

\subsection*{Acknowledgements}

R.F.R. carried out exploratory studies of the presented remote control
in the linear regime, supported by a UCSD CRES postdoctoral award.
R.F.R. coarse-grained the model of IVR from the literature, supported
by AFOSR award FA9550-18-1-0289. The development of the remote-control
model in the nonlinear regime via input-output theory by M.D. and
J.Y.-Z. was supported by U.S. Department of Energy, Office of Science,
Early Career Research Program under Award No. DE-SC0019188. We acknowledge Dr. Johannes Schachenmayer for discussions on nonlinear optical control of polaritons. M.D. thanks
Jorge Campos-Gonz\'alez-Angulo, Garret Wiesehan, Juan Perez-Sanchez,
and Luis Mart\'inez-Mart\'inez for useful discussions.

\subsection*{Author contributions}

M.D. designed, carried out, and analyzed the calculations. R.F.R.
guided the design and analysis of the calculations. J.Y.-Z. designed,
conceived, and supervised the project. M.D. wrote the paper with input
from all other authors.

\subsection*{Competing interests}

The authors declare no competing interests.

\subsection*{Additional information}

\textbf{Supplementary information} is available for this paper.\\
\textbf{Correspondence and requests for materials} should be addressed
to J.Y.-Z.

\bibliographystyle{apsrev4-1}
\bibliography{remote-control}

\clearpage

\onecolumngrid
\renewcommand{\figurename}{Supplementary Figure}
\renewcommand{\thesubsection}{\arabic{subsection}} 
\setcounter{figure}{0}
\setcounter{section}{0}
\setcounter{equation}{0}
\renewcommand{\theequation}{S\arabic{equation}} 
\setcounter{page}{1} \renewcommand{\thepage}{\arabic{page}}

\title{Supplementary Notes for ``Remote control of chemistry in optical
cavities'' }

\author{Matthew Du, Raphael F. Ribeiro, Joel Yuen-Zhou$^{*}$}

\address{Department of Chemistry and Biochemistry, University of California
San Diego, La Jolla, California 92093, United States}
\email{joelyuen@ucsd.edu}

\maketitle

\onecolumngrid

\subsection{Tc-glyoxylic acid as choice of `remote catalyst'\label{subsec:glyoxylic_acid}}

The choice of the Tc conformer of glyoxylic acid \cite{Olbert-Majkut2014cpl,Olbert-Majkut2014jpca}
as the `remote catalyst' (RC) is motivated by the following:
\begin{enumerate}
\item $|\text{RC}\rangle$ is an efficient absorber and should strongly
couple to light given sufficient RC concentration, 
\item the highest polariton of the device essentially has character of only
RC and its cavity (Supplementary Fig. \ref{fig:SI_rc_MF}b),
\item solvent vibrational degrees of freedom mediate efficient relaxation
between this highest polariton and RC dark states via one-phonon or
multiphonon processes (Supplementary Fig. \ref{fig:SI_rc_pump}),
\item $|\text{RC}\rangle$ is photochemically stable (\emph{i.e.}, does
not interconvert to conformers with different vibrational frequencies).
\item the lowest polariton of the device has significant character of R,
RC, and their host cavities.
\end{enumerate}
While excitation of the first overtone (second excited state) of Tc-glyoxylic
acid induces a number of conformational reactions \cite{Olbert-Majkut2014jpca,Olbert-Majkut2014cpl},
these side processes can be safely neglected for this work. The 1$\rightarrow$2
$|\text{RC}\rangle$ transition of Tc-glyoxylic acid is sufficiently
detuned from the 0 $\rightarrow$ 1 transitions of $|\text{R}\rangle$,
$|\text{RC}\rangle$, and the strongly coupled cavity photons, such
that the former transition contributes insignificantly to the polariton
dynamics of the device.

To the best of our knowledge, the OH stretch frequency is not reported
for Tc-glyoxylic acid in a Kr matrix, in which the fast-channel isomerization
from R to \textit{P} was observed \cite{Schanz2005}. We thus approximate
the frequency and anharmonic coupling constant of $|\text{RC}\rangle$
as averages of those for Tc-glyoxylic acid in Ar ($\omega_{10}=3473.5$
and $\Delta=-86$ $\text{cm}^{-1}$, respectively \cite{Olbert-Majkut2014jpca})
and Xe ($\omega_{10}=3436.5$ and $\Delta=-92$ $\text{cm}^{-1}$,
respectively, for the conformer located in a `tight site' of the Xe
matrix \cite{Olbert-Majkut2014cpl}). 

\subsection{Isomerization model\label{subsec:isomerization}}

In the main text, we considered the following mechanism for the fast
isomerization pathway from R to P in a Kr matrix \cite{Schanz2005}.
Proceeding laser excitation, $|\text{R}\rangle$ first transfers energy
via intramolecular vibrational redistribution (IVR) into $|\text{P}\rangle$,
specifically the lower eigenstate that results (primarily) from mixing
of the seventh localized torsional overtone of R and P \cite{Schanz2005}.
Then, $|\text{P}\rangle$ vibrationally relaxes into the localized
potential wells of either R or P. This mechanism resembles those first
proposed for the isomerization induced by pulsed \cite{Schanz2005}
or continuous-wave \cite{Hall1963} excitation. However, there is
an alternative mechanism where $|\text{R}\rangle$ couples to a few
discrete states that subsequently decay into continua, giving rise
to P \cite{Hamm2008}. Nevertheless, the qualitative structure of the model
is the same as the first one. Following the first mechanism
for pulse-induced isomerization \cite{Schanz2005}, we take $|\text{P}\rangle$
to be 40 cm$^{-1}$ lower in energy than $|\text{R}\rangle$. This
energy gap (illustrated by the difference between the minima of potential
energy curves 3 and 4 in Fig. 7 of \cite{Schanz2005}) was obtained
by adjusting the barrier height of the torsional double well to a
value ($3610\text{ cm}^{-1}$) that allows the potential energy curves
of $|\text{R}\rangle$and $|\text{P}\rangle$ to cross \cite{Schanz2005}. 

\subsection{Reaction efficiency in the absence of strong light-matter coupling\label{subsec:reaction} }

In this Supplementary Note, we calculate the efficiency of the fast
isomerization from \textit{cis}-HONO to \textit{trans}-HONO in the
bare case, \textit{i.e.}, in the absence of strong light-matter coupling.
We apply input-output theory \cite{Gardiner1985,Ciuti2006,SteckBook}
to the reaction model described in the main text and in Supplementary
Note \ref{subsec:isomerization}. IVR from $|\text{R}\rangle$ to
$|\text{P}\rangle$ is treated as a coupling between the two states,
while the other reaction steps (\textit{e.g.}, absorption, relaxation
into \textit{trans} torsional potential well) are assumed to occur
via coupling to a continuum of bath modes under the wide-band \cite{NitzanBook}
approximation. 

\begin{figure}
\includegraphics[width=0.85\textwidth]{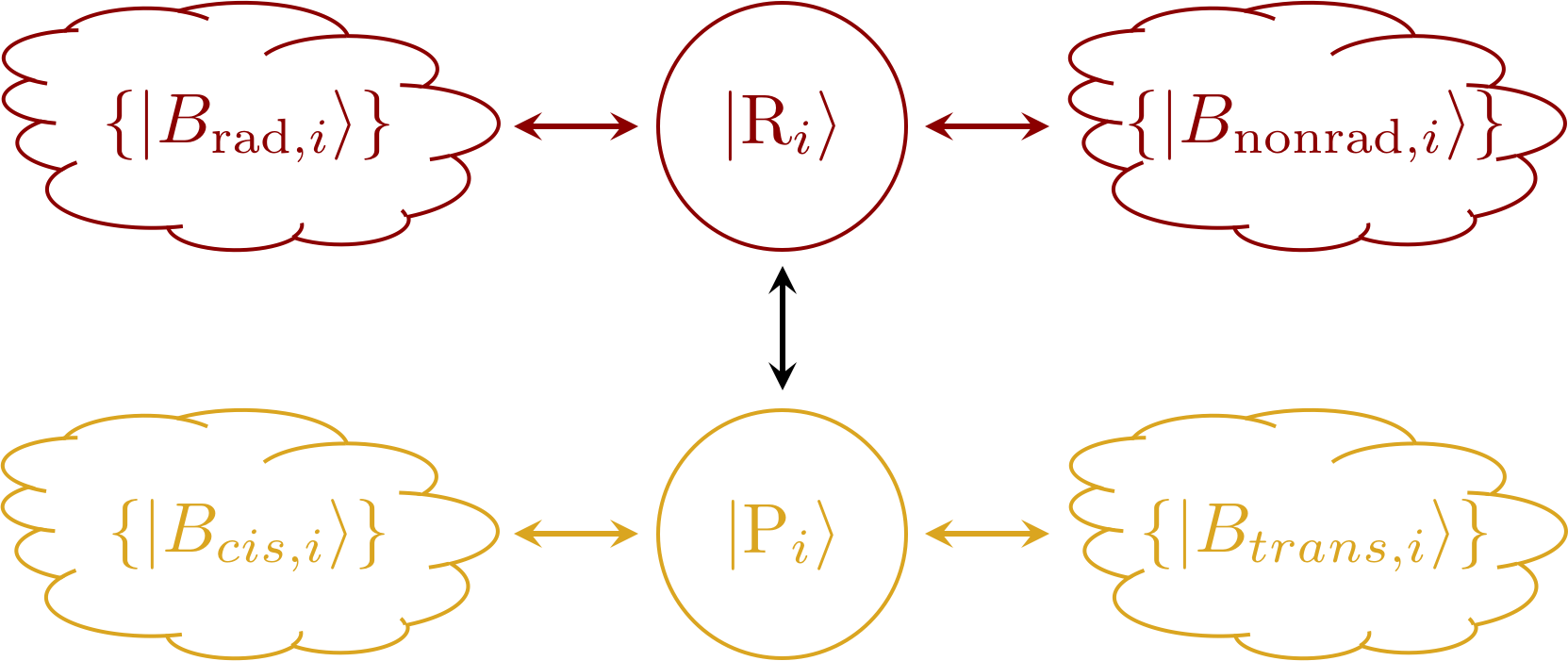}

\caption{\textbf{Schematic of the system-bath interactions of the R $\rightarrow$
P fast isomerization channel in the absence of strong light-matter
coupling.} See equation (\ref{eq:h0})) for the Hamiltonian $H_{0}$.
The $i$th R molecule has an OH stretch $|\text{R}_{i}\rangle$ (dark
red circle) which is coupled to a set ${|B_{\text{rad},i}\rangle}$
of radiative bath states (left dark red cloud) that contribute to
its excitation and a set ${|B_{\text{nonrad},i}\rangle}$ of nonradiative
bath states (right dark red cloud) that contribute to its decay. $|\text{R}_{i}\rangle$
is coupled to the torsional overtone state $|\text{P}_{i}\rangle$
(yellow circle), which can decay into either the $cis$ or $trans$
potential wells of HONO via coupling to ${|B_{cis,i}\rangle}$ (left
yellow cloud) or ${|B_{trans,i}\rangle}$ (right yellow cloud), respectively.
\label{fig:SI_io_bare}}
\end{figure}

Understanding that photonic bath modes induce excitation but nonradiative
bath modes are the main contributors to relaxation of $|\text{R}\rangle$
\cite{NitzanBook}, we can write the Hamiltonian as we did for the
polaritonic device (Methods): $H_{0}=H_{\text{system},0}+H_{\text{bath},0}+H_{\text{system}-\text{bath},0}$.
Specifically (hereafter, $\hbar=1$), \begin{subequations}\label{eq:h0}
\begin{align}
H_{\text{system},0} & =H_{\text{R}}+H_{\text{P}}+H_{\text{R}-\text{P}},\label{eq:hS0}\\
H_{\text{bath},0} & =\sum_{x=\text{rad},\text{nonrad}}\sum_{i=1}^{N_{\text{R}}}\int_{-\infty}^{\infty}d\omega\,\omega B_{x,i}^{\dagger}(\omega)B_{x,i}(\omega)+\sum_{y=cis,trans}\sum_{i=1}^{N_{\text{R}}}\int_{-\infty}^{\infty}d\omega\,\omega|B_{y,i}(\omega)\rangle\langle B_{y,i}(\omega)|,\label{eq:hB0}\\
H_{\text{system}-\text{bath},0} & =\sum_{x=\text{rad},\text{nonrad}}\frac{\sqrt{\gamma_{x}}}{\sqrt{2\pi}}\sum_{i=1}^{N_{\text{R}}}\int_{-\infty}^{\infty}d\omega[a_{\text{R},i}^{\dagger}B_{x,i}(\omega)+\text{h.c.}]+\sum_{y=cis,trans}\frac{\sqrt{\Gamma_{y}}}{\sqrt{2\pi}}\sum_{i=1}^{N_{\text{R}}}\int_{-\infty}^{\infty}d\omega[|\text{P}_{i}\rangle\langle B_{y,i}(\omega)|+\text{h.c.}],\label{eq:hSB0}
\end{align}
\end{subequations}where $H_{\text{R}}$, $H_{\text{P}}$, and $H_{\text{R}-\text{P}}$
are defined respectively in equations (\ref{eq:hR}), (\ref{eq:hP}),
and (\ref{eq:hRP}) of Methods. The various interactions between $|\text{R}_{i}\rangle=a_{\text{R},i}^{\dagger}|\text{G}\rangle$
, $|\text{P}_{i}\rangle$, and their baths are summarized in Fig.
\ref{fig:SI_io_bare}. $|\text{G}\rangle$ is the vacuum molecular
ground state. The symbols $a_{\text{R},i}^{\dagger},|\text{P}_{i}\rangle$
and their Hermitian conjugates are defined in Methods. The bath modes
are taken to be linearly coupled to $|\text{R}\rangle=\sum_{i=1}^{N_{\text{R}}}|\text{R}_{i}\rangle/\sqrt{N_{\text{R}}}$
and $|\text{P}\rangle=\sum_{i=1}^{N_{\text{R}}}|\text{P}_{i}\rangle/\sqrt{N_{\text{R}}}$
and are written in a form convenient for application of input-output
theory \cite{SteckBook}. In particular, the operator $B_{x,i}^{\dagger}(\omega)$
($B_{x,i}(\omega)$) creates (annihilates) an $|\text{R}\rangle$
bath mode of type $x$ and frequency $\omega$ on R molecule $i$
and satisfies the bosonic commutation relations $[B_{x,i}(\omega),B_{x',i'}^{\dagger}(\omega')]=\delta_{x,x'}\delta_{i,i'}\delta(\omega-\omega')$
and $[B_{x,i}(\omega),B_{x',i'}(\omega')]=0$, where $\delta_{x,x'},\delta_{i,i'}$
are Kronecker deltas and $\delta(\cdot)$ is the Dirac delta function.
The label $x=\text{rad}$ ($x=\text{nonrad}$) indicates a radiative
(nonradiative) bath mode that excites (relaxes) R due to coupling
to the OH stretch mode $|\text{R}\rangle$ with strength $\gamma_{\text{rad}}$
($\gamma_{\text{nonrad}}$). Similarly, the state $|B_{y,i}(\omega)\rangle$,
representing a $|\text{P}\rangle$ bath mode of type $y$ and frequency
$\omega$, satisfies the relation $\langle B_{y,i}(\omega)|B_{y',i'}(\omega')\rangle=\delta_{y,y'}\delta_{i,i'}\delta(\omega-\omega')$.
The label $y=cis$ ($y=trans$) indicates a bath that relaxes $|\text{P}\rangle$
into the \textit{cis} (\textit{trans}) well via coupling to $|\text{P}\rangle$
with strength $\Gamma_{cis}$ ($\Gamma_{trans}$). 

We use state representation for $|\text{P}\rangle$ in equations (\ref{eq:hB0})
and (\ref{eq:hSB0}) (as well as equations (\ref{eq:hP}) and (\ref{eq:hRP})
of Methods) because this state is the seventh overtone of the torsional
coordinate, \textit{i.e.}, $|\text{P}_{i}\rangle=\sum_{m=1}^{\infty}[c_{cis,m}(a_{\tau,cis,i}^{\dagger})^{m}/\sqrt{m!}+c_{trans,m}(a_{\tau,trans,i}^{\dagger})^{m}/\sqrt{m!}]|\text{G}\rangle$.
The creation (annihilation) operator $a_{\tau,x,i}^{\dagger}$ ($a_{\tau,x,i}$)
is associated with the diabatic torsional state $x=cis,trans$ for
the $i$th HONO molecule . The $m$th state of this mode has expansion
coefficient $c_{x.m}$. If $|\text{P}\rangle$ overlaps most with
the seventh localized overtones of each conformational isomer, we
can write $|\text{P}_{i}\rangle\approx[c_{cis,8}(a_{\tau,cis,i}^{\dagger})^{8}/\sqrt{8!}+c_{trans,8}(a_{\tau,trans,i}^{\dagger})^{8}/\sqrt{8!}]|\text{G}\rangle$.
Likewise, the use of state representation for the bath modes of $|\text{P}\rangle$
is motivated by the fact they are also likely to be combination/overtone
modes, given that they must be high in energy to efficiently couple
to the torsional state and induce relaxation.

We next use input-output theory to calculate the reaction efficiency.
For notational convenience, we define $\beta_{x}(\omega)=\sum_{i=1}^{N_{\text{R}}}B_{x,i}(\omega)/\sqrt{N_{\text{R}}}$
and $\beta_{y}(\omega)=\sum_{i=1}^{N_{\text{R}}}|\text{G}\rangle\langle B_{y,i}(\omega)|/\sqrt{N_{\text{R}}}$
, and use $\frac{d\mathcal{O}(t)}{dt}=-i[\mathcal{O},H_{0}]$ for
any operator $\mathcal{O}$ which is time-independent in the Schr\"odinger
picture. Then we obtain the following Heisenberg equations of motion:
\begin{align}
\frac{dP_{\text{R}}(t)}{dt} & =-i\omega_{\text{R}}P_{\text{R}}(t)-iV_{\text{R}-\text{P}}P_{\text{P}}(t)-i\sum_{x=\text{rad},\text{nonrad}}\frac{\sqrt{\gamma_{x}}}{\sqrt{2\pi}}\int_{-\infty}^{\infty}d\omega\beta_{x}(\omega)(t),\label{eq:pR1b4}\\
\frac{dP_{\text{P}}(t)}{dt} & =-i\omega_{\text{P}}P_{\text{P}}(t)-iV_{\text{R}-\text{P}}P_{\text{R}}(t)-i\sum_{y=cis,trans}\frac{\sqrt{\Gamma_{y}}}{\sqrt{2\pi}}\int_{-\infty}^{\infty}d\omega\beta_{y}(\omega)(t),\label{eq:pPb4}\\
\frac{d\beta_{x}(\omega)(t)}{dt} & =-i\omega\beta_{x}(\omega)(t)-i\frac{\sqrt{\gamma_{x}}}{\sqrt{2\pi}}P_{\text{R}}(t),\\
\frac{d\beta_{y}(\omega)(t)}{dt} & =-i\omega\beta_{y}(\omega)(t)-i\frac{\sqrt{\Gamma_{y}}}{\sqrt{2\pi}}P_{\text{P}}(t),
\end{align}
for polarization operators $P_{\text{R}}=\sum_{i=1}^{N_{\text{R}}}a_{\text{R},i}/\sqrt{N_{\text{R}}}$
and $P_{\text{P}}=\sum_{i=1}^{N_{\text{R}}}|\text{G}\rangle\langle\text{P}_{i}|/\sqrt{N_{\text{R}}}$
. Notice that 
\begin{align}
\frac{d[\beta_{x}(\omega)(t)e^{i\omega t}]}{dt} & =-i\frac{\sqrt{\gamma_{x}}}{\sqrt{2\pi}}P_{\text{R}}(t)e^{i\omega t},\label{eq:bxe}\\
\frac{d[\beta_{y}(\omega)(t)e^{i\omega t}]}{dt} & =-i\frac{\sqrt{\Gamma_{y}}}{\sqrt{2\pi}}P_{\text{P}}(t)e^{i\omega t}.\label{eq:bye}
\end{align}
Defining $t_{\text{in}}<t$ and $t_{\text{out}}>t$, we integrate
equation (\ref{eq:bxe}) and obtain the relations\begin{subequations}\label{eq:binout}
\begin{align}
\beta_{x}(\omega)(t) & =\beta_{x}(\omega)(t_{\text{in}})e^{-i\omega(t-t_{\text{in}})}-i\frac{\sqrt{\gamma_{x}}}{\sqrt{2\pi}}\int_{t_{\text{in}}}^{t}dt'P_{\text{R}}(t')e^{-i\omega(t-t')},\label{eq:intx}\\
\beta_{x}(\omega)(t) & =\beta_{x}(\omega)(t_{\text{out}})e^{-i\omega(t-t_{\text{out}})}+i\frac{\sqrt{\gamma_{x}}}{\sqrt{2\pi}}\int_{t}^{t_{\text{out}}}dt'P_{\text{R}}(t')e^{-i\omega(t-t')}.\label{eq:intxo}
\end{align}
\end{subequations}Plugging in equation (\ref{eq:binout}) for $x=\text{rad}$
into (\ref{eq:pR1b4}) yields \cite{SteckBook}\begin{subequations}\label{eq:pinout}
\begin{align}
\frac{dP_{\text{R}}(t)}{dt} & =-i\omega_{\text{R}}P_{\text{R},1}(t)-iV_{\text{R}-\text{P}}P_{\text{P}}(t)-i\frac{\sqrt{\gamma_{\text{nonrad}}}}{\sqrt{2\pi}}\int_{-\infty}^{\infty}d\omega\beta_{\text{nonrad}}(\omega)(t)-\sqrt{\gamma_{\text{rad}}}P_{\text{R},\text{in},\text{rad}}(t)-(\gamma_{\text{rad}}/2)P_{\text{R}}(t),\label{eq:pioin}\\
\frac{dP_{\text{R}}(t)}{dt} & =-i\omega_{\text{R}}P_{\text{R},1}(t)-iV_{\text{R}-\text{P}}P_{\text{P}}(t)-i\frac{\sqrt{\gamma_{\text{nonrad}}}}{\sqrt{2\pi}}\int_{-\infty}^{\infty}d\omega\beta_{\text{nonrad}}(\omega)(t)-\sqrt{\gamma_{\text{rad}}}P_{\text{R},\text{out},\text{rad}}(t)+(\gamma_{\text{rad}}/2)P_{\text{R}}(t),\label{eq:pioout}
\end{align}
\end{subequations}where the input and output polarizations are
\begin{align}
P_{\text{R},\text{in},z}(t) & =\frac{i}{\sqrt{2\pi}}\int_{-\infty}^{\infty}d\omega\beta_{z}(\omega)(t_{\text{in}})e^{-i\omega(t-t_{\text{in}})},\\
P_{\text{R},\text{out},z}(t) & =\frac{i}{\sqrt{2\pi}}\int_{-\infty}^{\infty}d\omega\beta_{z}(\omega)(t_{\text{out}})e^{-i\omega(t-t_{\text{out}})},
\end{align}
respectively, for $z=\text{rad},\text{nonrad},cis,trans$. Subtracting
the second from the first row of equation (\ref{eq:pinout}) gives
\begin{equation}
P_{\text{R},\text{out},\text{rad}}(t)-P_{\text{R},\text{in},\text{rad}}(t)=\sqrt{\gamma_{\text{rad}}}P_{\text{R}}(t).\label{eq:iorad}
\end{equation}
This is an example of \emph{input-output relation}. We analogously
obtain relations for the $x=\text{nonrad}$ and $y=cis,trans$ baths,
\begin{align}
P_{\text{R},\text{out},\text{nonrad}}(t)-P_{\text{R},\text{in},\text{nonrad}}(t) & =\sqrt{\gamma_{\text{nonrad}}}P_{\text{R}}(t),\label{eq:io0R}\\
P_{\text{P},\text{out},y}(t)-P_{\text{P},\text{in},y}(t) & =\sqrt{\Gamma_{y}}P_{\text{P}}(t).\label{eq:io0P}
\end{align}
Equations (\ref{eq:io0R}) and (\ref{eq:io0P}) will prove useful
below.

We want to express the output polarizations above in terms of $P_{\text{R},\text{in},\text{rad}}$.
To proceed, we first substitute equation (\ref{eq:intx}) for $x=\text{nonrad}$
into (\ref{eq:pioin}) \cite{SteckBook},

\begin{align}
\frac{dP_{\text{R}}(t)}{dt} & =-i\omega_{\text{R}}P_{\text{R},1}(t)-iV_{\text{R}-\text{P}}P_{\text{P}}(t)-\sum_{x=\text{rad},\text{nonrad}}\sqrt{\gamma_{x}}P_{\text{R},\text{in},x}(t)-\sum_{x=\text{rad},\text{nonrad}}(\gamma_{x}/2)P_{\text{R}}(t).\label{eq:pR1close}
\end{align}
The analogous equation for $P_{\text{P}}$ is
\begin{align}
\frac{dP_{\text{P}}(t)}{dt} & =-i\omega_{\text{P}}P_{\text{P}}(t)-iV_{\text{R}-\text{P}}P_{\text{R},1}(t)-\sum_{y=cis,trans}\sqrt{\Gamma_{y}}P_{\text{P},\text{in},y}(t)-\sum_{y=cis,trans}(\Gamma_{y}/2)P_{\text{P}}(t).\label{eq:pPclose}
\end{align}
Recalling that $|\text{R}\rangle$ is only excited by the photonic
bath via \textit{weak} light-matter coupling and $|\text{P}\rangle$
is only excited by IVR coupling to $|\text{R}\rangle$, we hereafter
take $P_{\text{R},1,\text{in},\text{nonrad}}(t),P_{\text{P},\text{in},y}(t)=0$
and $\gamma_{\text{rad}}\ll\gamma_{\text{nonrad}}$ to arrive at the
Heisenberg-Langevin equations:

\begin{align}
\frac{dP_{\text{R}}(t)}{dt} & =-i(\omega_{\text{R}}-i\gamma_{\text{nonrad}}/2)P_{\text{R}}(t)-iV_{\text{R}-\text{P}}P_{\text{P}}(t)-\sqrt{\gamma_{\text{rad}}}P_{\text{R},\text{in},\text{rad}}(t)\label{eq:pR1done}\\
\frac{dP_{\text{P}}(t)}{dt} & =-i[\omega_{\text{P}}-i(\Gamma_{cis}+\Gamma_{trans})/2]P_{\text{P}}(t)-iV_{\text{R}-\text{P}}P_{\text{R}}(t).\label{eq:pPdone}
\end{align}
We now solve for $P_{\text{R}}$ and $P_{\text{P}}$ in the frequency
domain to calculate the (frequency-resolved) absorption of R and the
reaction efficiency. Taking the Fourier transform $\mathcal{F}[f(t)]=\int_{-\infty}^{\infty}dt\,e^{i\omega t}f(t)$
of equations (\ref{eq:pR1done}) and (\ref{eq:pPdone}), we obtain
\begin{align}
P_{\text{R}}(\omega) & =S_{\text{R},0}(\omega)P_{\text{R},\text{in},\text{rad}}(\omega),\label{eq:pR1c}\\
P_{\text{P}}(\omega) & =S_{\text{P},0}(\omega)P_{\text{R},\text{in},\text{rad}}(\omega),\label{eq:pPc}
\end{align}
where
\begin{align}
S_{\text{R},0}(\omega) & =\frac{-i\sqrt{\gamma_{\text{rad}}}[\omega-\omega_{\text{P}}+i(\Gamma_{cis}+\Gamma_{trans})/2]}{(\omega-\omega_{\text{R}}+i\gamma_{\text{nonrad}}/2)[\omega-\omega_{\text{P}}+i(\Gamma_{cis}+\Gamma_{trans})/2]-V_{\text{R}-\text{P}}^{2}},\label{eq:cR0b4}\\
S_{\text{P},0}(\omega) & =S_{\text{R},0}(\omega)\frac{V_{\text{R}-\text{P}}}{\omega-\omega_{\text{P}}+i(\Gamma_{cis}+\Gamma_{trans})/2}.\label{eq:cP0b4}
\end{align}

We proceed to calculate the absorption of R and the reaction efficiency.
The steady-state absorbance of R, \textit{i.e.}, the fraction of the
input energy that is dissipated by the nonradiative bath coupled to
R, is\begin{subequations}\label{eq:aR0b4}
\begin{align}
\text{absorbance}_{\text{R},0}(\omega) & =\frac{\langle P_{\text{R},\text{out},\text{R}}^{\dagger}(\omega)P_{\text{R},\text{out},\text{R}}(\omega)\rangle}{\langle P_{\text{R},\text{in},\text{LM}}^{\dagger}(\omega)P_{\text{R},\text{in},\text{LM}}(\omega)\rangle}\\
 & =\gamma_{\text{nonrad}}|S_{\text{R},0}(\omega)|^{2}\label{eq:gRcb4}\\
 & =\frac{\gamma_{\text{rad}}\gamma_{\text{nonrad}}}{\left|\omega-\omega_{\text{R}}+i\gamma_{\text{nonrad}}/2-\frac{V_{\text{R}-\text{P}}^{2}}{\omega-\omega_{\text{P}}+i(\Gamma_{cis}+\Gamma_{trans})/2}\right|^{2}}\label{eq:gRcb4expl}
\end{align}
\end{subequations}where we have used input-output equation (\ref{eq:io0R}),
as well as equations (\ref{eq:pR1c}), and (\ref{eq:cR0b4}). In agreement
with physical intuition, equation (\ref{eq:gRcb4expl}) says that
the absorption of R peaks near $\omega_{\text{R}}$ but is slightly
offset by an IVR-induced energy correction represented by the fraction
in the denominator. Analogous to the absorbance of R, the steady-state
reaction efficiency $\eta_{0}$ is the fraction of the input energy
whose bath-induced dissipation relaxes $|\text{P}\rangle$ into the
\textit{trans} localized potential well. Thus, \begin{subequations}\label{eq:eta0b4}
\begin{align}
\eta_{0}(\omega) & =\frac{\langle P_{\text{P},\text{out},trans}^{\dagger}(\omega)P_{\text{P},\text{out},trans}(\omega)\rangle}{\langle P_{\text{R},\text{in},\text{rad}}^{\dagger}(\omega)P_{\text{R},\text{in},\text{rad}}(\omega)\rangle}\\
 & =\Gamma_{trans}|S_{\text{P},0}(\omega)|^{2}\\
 & =\frac{\text{absorbance}_{\text{R},0}(\omega)}{\gamma_{\text{nonrad}}}\gamma_{\text{R}\rightarrow\text{P}}(\omega).\label{eq:feta0}
\end{align}
\end{subequations}We have used input-output equation (\ref{eq:io0P}),
as well as equations (\ref{eq:pPc}), (\ref{eq:cP0b4}), and (\ref{eq:gRcb4}).
We also recognized that 
\begin{equation}
\gamma_{\text{R}\rightarrow\text{P}}(\omega)=\frac{V_{\text{R}-\text{P}}^{2}\Gamma_{trans}}{(\omega-\omega_{\text{P}})^{2}+[(\Gamma_{cis}+\Gamma_{trans})/2]^{2}}\label{eq:gRP}
\end{equation}
is exactly the (quantum mechanical) steady-state rate of transition
from the (energy-broadened) $|\text{R}\rangle$ state of frequency
$\omega$ into the \textit{trans} well \cite{Segal2001,NitzanBook}.
Then the bare reaction efficiency can be intuitively expressed as
\begin{equation}
\eta_{0}(\omega)=\text{absorbance}_{\text{R},0}(\omega)\text{QY}_{\text{R}\rightarrow\text{P}}(\omega),\label{eq:eta0intu}
\end{equation}
where $\text{QY}_{\text{R}\rightarrow\text{P}}=\gamma_{\text{R}\rightarrow\text{P}}(\omega)/\gamma_{\text{nonrad}}$
is an isomerization quantum yield, though defined differently than
in the experimental report of fast-channel HONO isomerization \cite{Schanz2005}.
Satisfyingly, equation (\ref{eq:eta0intu}) says that the reaction
efficiency depends on how well R first absorbs at $\omega$ into $|\text{R}\rangle$
and then isomerizes by IVR coupling into $|\text{P}\rangle$ at this
same frequency. In the simulations shown in Figs. \ref{fig:rc_ON}
and \ref{fig:rc_g} of the main text, we use (see equation (\ref{eq:eta0mt})
in Methods) $\eta_{0}(\omega_{\text{R}})\approx0.0023$, calculated
from equation (\ref{eq:feta0}) with the previously reported $\text{absorbance}_{\text{R},0}(\omega_{\text{R}})\approx0.07$
\cite{Schanz2005}, absorption linewidth $\gamma_{\text{nonrad}}\approx5\text{ cm}^{-1}$
\cite{Schanz2005}, and isomerization rate $\gamma_{\text{R}\rightarrow\text{P}}(\omega_{\text{R}})\approx0.167\text{ cm}^{-1}$
(or $5\text{ ns}^{-1}$) \cite{Schanz2005}.

For $\omega\neq\omega_{\text{R}}$, computing $\eta_{0}(\omega)$
requires (see equation (\ref{eq:feta0})) explicit evaluation of $\gamma_{\text{R}\rightarrow\text{P}}(\omega)$,
equation (\ref{eq:gRP}), and $\text{absorption}_{\text{R},0}(\omega)$,
equation (\ref{eq:gRcb4expl}). For calculating $\gamma_{\text{R}\rightarrow\text{P}}(\omega)$,
in both this and polaritonic (Supplementary Note \ref{subsec:reactionSC})
cases, we assign values to $\Gamma_{cis}+\Gamma_{trans}$ and $V_{\text{R}-\text{P}}^{2}\Gamma_{trans}$.
First, we set $\Gamma_{cis}+\Gamma_{trans}$ to the rate of $|\text{P}\rangle$
population relaxation. Treating the process as one-phonon emission
of the diabatic eighth-excited torsional state of \textit{cis}-HONO
in the harmonic limit, $\Gamma_{cis}+\Gamma_{trans}$ is 8 times the
population relaxation rate of the singly excited torsion of \textit{cis}-HONO
\cite{NitzanBook}. We make this harmonic approximation because the
population decay rate of $|\text{P}\rangle$ is not reported (to the
best of our knowledge), but that of the $v=1$ torsional state of
\textit{cis}-HONO can be estimated. Specifically, we assume that this
single excitation relaxes at the same rate ($1.67\text{ cm}^{-1}$
, or $(20\text{ ps})^{-1}$ \cite{Schanz2005}) as $|\text{R}\rangle$,
in analogy to the similarity of their absorption linewidths \cite{Khriachtchev2000}.
While anharmonicity, other relaxation channels (\textit{e.g.}, multiphonon
processes), and pure dephasing may significantly contribute to the
decay of $|\text{P}\rangle$, these contributions are not well-characterized
(to the best of our knowledge). Furthermore, modeling these contributions
is difficult and should not change the main conclusions of this work.
Having set $\Gamma_{cis}+\Gamma_{trans}=8\times1.67\text{ cm}^{-1}$,
we assign $V^{2}\Gamma_{trans}\approx275\text{ cm}^{-3}$ to afford
the experimentally observed value $\gamma_{\text{R}\rightarrow\text{P}}(\omega_{\text{R}})=0.167\text{ cm}^{-1}$
\cite{Schanz2005} from equation (\ref{eq:gRP}). 

Like $\gamma_{\text{R}\rightarrow\text{P}}(\omega)$, explicit calculation
of $\text{absorption}_{\text{R},0}(\omega)$ using equation (\ref{eq:gRcb4expl})
requires an unknown (to the best of our knowledge) parameter, namely
$V_{\text{R}-\text{P}}$. Since the reported \cite{Schanz2005} $V_{\text{R}-\text{P}}\approx15\text{ cm}^{-1}$
is incorrect \cite{Hamm2008}, we evaluate $\text{absorbance}_{\text{R},0}(\omega)$
(equation (\ref{eq:gRcb4expl})) by assuming $V_{\text{R}-\text{P}}$
is small, 
\begin{equation}
\text{absorbance}_{\text{R},0}(\omega)\approx\frac{\gamma_{\text{rad}}\gamma_{\text{nonrad}}}{(\omega-\omega_{\text{R}})^{2}+(\gamma_{\text{nonrad}}/2)^{2}},\label{eq:small}
\end{equation}
which intuitively is the Lorentzian lineshape for $|\text{R}\rangle$
in the absence of IVR. The criterion to justify equation (\ref{eq:small})
can be derived as follows. From equations (\ref{eq:gRcb4}) and (\ref{eq:cR0b4}),
the poles of the absorption are given by 
\begin{equation}
\omega_{\text{average}}\pm\frac{\sqrt{\omega_{\text{squared difference}}}}{2},\label{eq:peak}
\end{equation}
as well as their complex conjugates. The frequency
\[
\omega_{\text{average}}=\frac{\omega_{\text{R}}+\omega_{\text{P}}-i(\gamma_{\text{nonrad}}+\Gamma_{cis}+\Gamma_{trans})/2}{2}
\]
is the average of the complex-valued energies $\omega_{\text{R}}-i\gamma_{\text{nonrad}}/2$
and $\omega_{\text{P}}+i(\Gamma_{cis}+\Gamma_{trans})/2$ of $|\text{R}\rangle$
and $|\text{P}\rangle$, respectively. When $V_{\text{R}-\text{P}}=0$,
the squared difference of these energies is
\[
\omega_{\text{squared difference}}=(\omega_{\text{R}}-\omega_{\text{P}})^{2}-(\gamma_{\text{nonrad}}-\Gamma_{cis}-\Gamma_{trans})^{2}/4-i(\omega_{\text{R}}-\omega_{\text{P}})(\gamma_{\text{nonrad}}-\Gamma_{cis}-\Gamma_{trans})-V_{\text{R}-\text{P}}^{2}.
\]
By analogy with the criteria for strong interaction \cite{Torma2015},
the IVR energy correction in equation (\ref{eq:gRcb4expl}) can be
neglected if the square of the IVR coupling is less than the `linewidth'
of $\omega_{\text{squared difference}}$:
\begin{equation}
V^{2}<|(\omega_{\text{R}}-\omega_{\text{P}})(\gamma_{\text{nonrad}}-\Gamma_{cis}-\Gamma_{trans})|.\label{eq:critw}
\end{equation}
Using the values for $\Gamma_{cis}+\Gamma_{trans}$ and $V^{2}\Gamma_{trans}$
from the previous paragraph, this inequality is satisfied if the probability
$\Gamma_{trans}/(\Gamma_{cis}+\Gamma_{trans})$ of decaying into the
\textit{trans} potential well is greater than $\approx0.06$. This
range of probabilities is reasonable, given that the measured isomerization
quantum yield (as defined in \cite{Schanz2005}) is 10\% \cite{Schanz2005}.

\subsection{Reaction efficiency for polaritonic device\label{subsec:reactionSC}}

In this Supplementary Note, we outline the derivation of the transient
spectra and reaction efficiency for the proposed polaritonic device
using input-output theory for pump-probe spectroscopy of vibrational
polaritons \cite{Ribeiro2018vp}. Besides the subtleties introduced
by the inclusion of nonlinear effects, this derivation follows the
same steps as those of the bare case (Supplementary Note \ref{subsec:reaction}).
In fact, the notation here is identical to that both Supplementary
Note \ref{subsec:reaction} and Methods, unless otherwise noted. 

\begin{figure}
\includegraphics[width=0.85\textwidth]{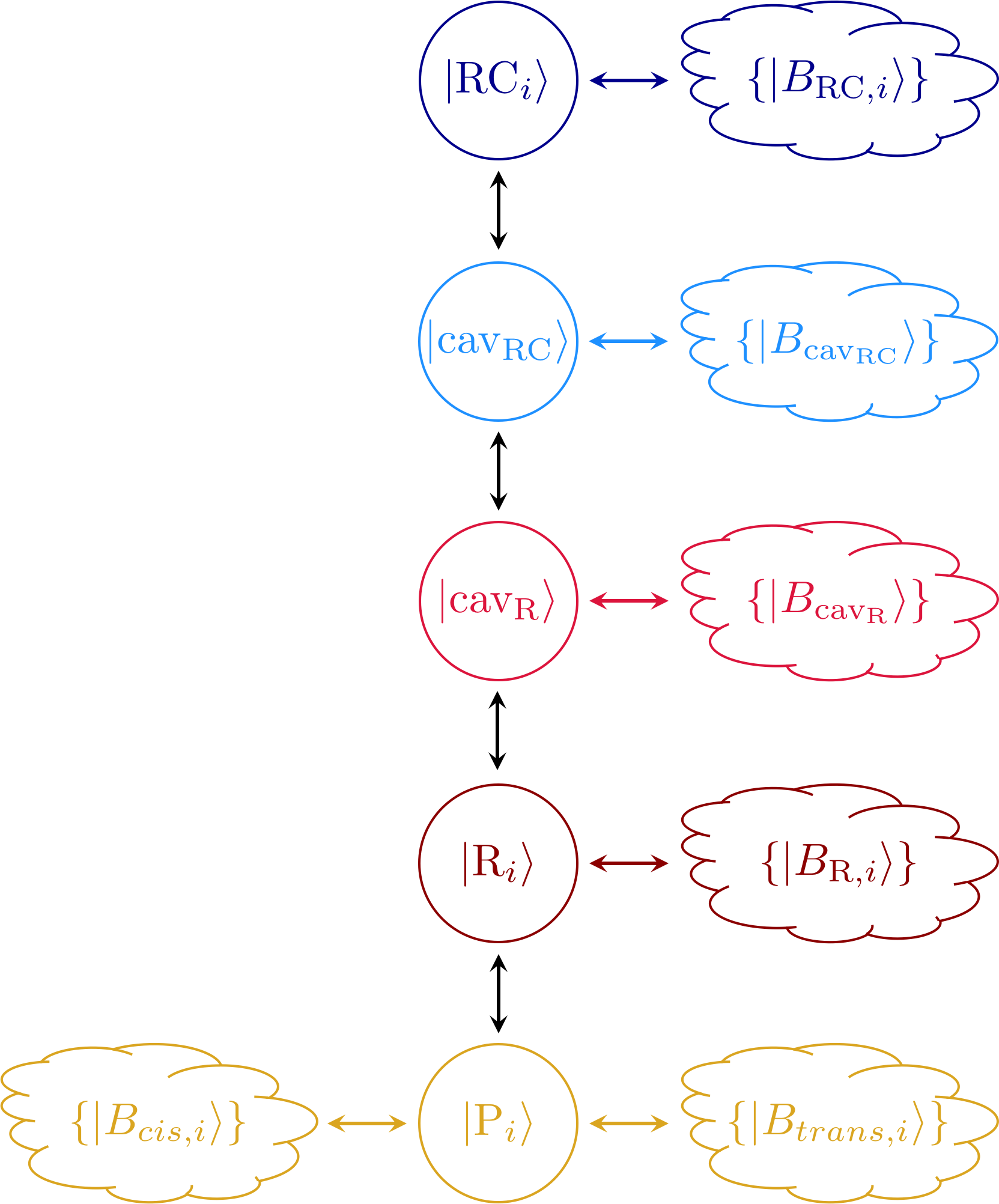}

\caption{\textbf{Schematic of the system-bath interactions of the R $\rightarrow$
P fast isomerization channel for the polaritonic device.} See equations
(\ref{eq:hS}), (\ref{eq:hBSB}), and those defining the terms therein,
for Hamiltonian $H$. The $i$th molecule of compound $x=\text{RC},\text{R}$
has an OH stretch excitation $|x_{i}\rangle$ (dark blue, dark red
circle), which is coupled to a set ${|B_{x}\rangle}$ of nonradiative
bath states (dark blue, dark red clouds) that contributes to its decay.
$|x_{i}\rangle$ is also coupled to photon $|\text{cav}_{x}\rangle$
of the cavity containing compound $x$. $|\text{cav}_{x}\rangle$
interacts with a set ${|B_{\text{cav}_{x}}\rangle}$ of radiative
bath states, as well as with the other cavity state. For probe excitation
of the polaritonic device, the states in ${|B_{\text{cav}_{\text{RC}}}\rangle}$
(light blue cloud) contribute to decay\textemdash in the form of transmission\textemdash of
$|\text{cav}_{\text{R}C}\rangle$ (light blue circle), while the states
in ${|B_{\text{cav}_{\text{R}}}\rangle}$ (light red cloud) contribute
to decay\textemdash in the form of reflection\textemdash and excitation
of $|\text{cav}_{\text{R}}\rangle$ (light red circle). Finally, $|\text{R}_{i}\rangle$
is coupled to another state, the torsional overtone $|\text{P}_{i}\rangle$
(yellow), which can concomitantly decay into either the $cis$ or
$trans$ potential wells of HONO via coupling to ${|B_{cis,i}\rangle}$
or ${|B_{trans,i}\rangle}$, respectively. \label{fig:SI_io_rc}}
\end{figure}

As discussed in Methods, the Hamiltonian for the polaritonic device
is $H=H_{\text{system}}+H_{\text{bath}}+H_{\text{system}-\text{bath}}$.
The first term on the right-hand side is given by equation (\ref{eq:hS}),
as well as equation (\ref{eq:hRCall}) in Methods. The latter two
terms read\begin{subequations}\label{eq:hBSB}
\begin{align}
H_{\text{bath}} & =\sum_{x=\text{RC},\,\text{cav}_{\text{RC}},\,\text{cav}_{\text{R}},\,\text{R},\,\text{P}}H_{\text{bath},x},\\
H_{\text{system}-\text{bath}} & =\sum_{x=\text{RC},\,\text{cav}_{\text{RC}},\,\text{cav}_{\text{R}},\,\text{R},\,\text{P}}H_{\text{system}-\text{bath},x}.
\end{align}
\end{subequations}In particular, \begin{subequations}\label{eq:hB}
\begin{align}
H_{\text{bath},x} & =\sum_{i=1}^{N_{x}}\int_{-\infty}^{\infty}d\omega\,\omega B_{x,i}^{\dagger}(\omega)B_{x,i}(\omega),\label{eq:hBRCR}\\
H_{\text{bath},\text{cav}_{x}} & =\int_{-\infty}^{\infty}d\omega\,\omega B_{\text{cav}_{x}}^{\dagger}(\omega)B_{\text{cav}_{x}}(\omega),\label{eq:hBcav}\\
H_{\text{bath},\text{P}} & =\sum_{y=cis,trans}\sum_{i=1}^{N_{\text{R}}}\int_{-\infty}^{\infty}d\omega\,\omega|B_{y,i}(\omega)\rangle\langle B_{y,i}(\omega)|,\label{eq:hBP}
\end{align}
\end{subequations}and\begin{subequations}\label{eq:hSB} 

\begin{align}
H_{\text{system}-\text{bath},x} & =\frac{\sqrt{\gamma_{x}}}{\sqrt{2\pi}}\sum_{i=1}^{N_{z}}\int_{-\infty}^{\infty}d\omega[a_{x,i}^{\dagger}B_{x,i}(\omega)+\text{h.c.}],\label{eq:hSBRCR}\\
H_{\text{system}-\text{bath},\text{cav}_{x}} & =\frac{\sqrt{\kappa_{x}}}{\sqrt{2\pi}}\int_{-\infty}^{\infty}d\omega[c_{x}^{\dagger}B_{\text{cav}_{x}}(\omega)+\text{h.c.}],\label{eq:hSBcav}\\
H_{\text{system}-\text{bath},\text{P}} & =\sum_{y=cis,trans}\frac{\sqrt{\Gamma_{y}}}{\sqrt{2\pi}}\sum_{i=1}^{N_{\text{R}}}\int_{-\infty}^{\infty}d\omega[|\text{P}_{i}\rangle\langle B_{y,i}(\omega)|+\text{h.c.}]\label{eq:hSBP}
\end{align}
\end{subequations}for $x=\text{RC},\text{R}$. The various interactions
between $|x_{i}\rangle=a_{x,i}^{\dagger}|\text{G}\rangle$, $|\text{cav}_{x}\rangle$,
$|\text{P}_{i}\rangle$, and their baths are summarized in Fig. \ref{fig:SI_io_rc}.
In this Supplementary Note, $|\text{G}\rangle$ is the molecular and
photonic vacuum ground state. The system-bath coupling parameters
$\gamma_{x},\kappa_{x}$ are defined in Methods. Note that the bath
modes for $|\text{R}\rangle$ only include the nonradiative degrees
of freedom that induce its decay (cf. equations (\ref{eq:hB0}) and
(\ref{eq:hBRCR})). Since probe excitation of the polaritonic system
would be carried out experimentally via laser impingement on the R
cavity mirror, the radiative input is considered to couple only to
the R cavity \cite{Ciuti2006,Ribeiro2018vp}. The operators $a_{\text{RC},i}^{\dagger},a_{\text{RC},i}$
and $c_{x}^{\dagger},c_{x}$ for RC and the cavities, respectively,
are defined in Methods. The operator $B_{\text{RC},i}^{\dagger}(\omega)$
($B_{\text{RC},i}(\omega)$) creates (annihilates) an $|\text{RC}\rangle=\sum_{i=1}^{N_{\text{RC}}}|\text{RC}_{i}\rangle/\sqrt{N_{\text{RC}}}$
nonradiative bath mode of frequency $\omega$ on RC molecule $i$
and satisfies the bosonic commutation relations $[B_{\text{RC},i}(\omega),B_{\text{RC},i'}^{\dagger}(\omega')]=\delta_{i,i'}\delta(\omega-\omega')$
and $[B_{\text{RC},i}(\omega),B_{\text{RC},i'}(\omega')]=0$. Analogously,
$B_{\text{cav}_{x}}^{\dagger}(\omega)$ ($B_{\text{cav}_{x}}(\omega)$)
creates (annihilates) a $x$ cavity radiative bath mode of frequency
$\omega$ and satisfies $[B_{\text{cav}_{x}}(\omega),B_{\text{cav}_{x}}^{\dagger}(\omega')]=\delta(\omega-\omega')$
and $[B_{\text{cav}_{x}}(\omega),B_{\text{cav}_{x}}(\omega')]=0$. 

To derive equations of motion for $|\text{RC}\rangle$, RC cavity,
R cavity, $|\text{R}\rangle$, and $|\text{P}\rangle$, we first carry
out standard input-output theory, specifically the steps taken in
Supplementary Note \ref{subsec:reaction} to obtain equations of motion
(\ref{eq:pR1close}) and (\ref{eq:pPclose}) in the bare case. If
we then account for the assumption made above that only the R cavity
couples to an input field, we arrive at Heisenberg-Langevin equations
(\ref{eq:pRC1}), (\ref{eq:cRC}), (\ref{eq:cR}), (\ref{eq:pR}),
and (\ref{eq:pP}), respectively, of Methods; in equation (\ref{eq:cR}),
the term
\begin{equation}
c_{\text{R},\text{in}}(t)=\frac{i}{\sqrt{2\pi}}\int_{-\infty}^{\infty}d\omega B_{\text{cav}_{\text{R}}}(\omega)(t_{\text{in}})e^{-i\omega(t-t_{\text{in}})}\label{eq:crin}
\end{equation}
is the R cavity input field that excites the R cavity at time $t_{\text{in}}$.
Furthermore, applying this assumption and the steps invoked in \ref{subsec:reaction}
to obtain the bare input-output relations (equations (\ref{eq:iorad})-(\ref{eq:io0P}))
yields the polaritonic input-output relations:\begin{subequations}\label{eq:io}
\begin{align}
c_{\text{RC},\text{out}}(t) & =\sqrt{\kappa_{\text{RC}}}c_{\text{RC}}(t),\\
c_{\text{R},\text{out}}(t)-c_{\text{R},\text{in}}(t) & =\sqrt{\kappa_{\text{R}}}c_{\text{R}}(t),\\
P_{z,\text{out}}(t) & =\sqrt{\gamma_{z}}P_{z}(t),\label{eq:ioz}\\
P_{\text{P},\text{out},y}(t) & =\sqrt{\Gamma_{y}}P_{\text{P}}(t),\label{eq:ioP}
\end{align}
\end{subequations}where 
\begin{align}
c_{x,\text{out}}(t) & =\frac{i}{\sqrt{2\pi}}\int_{-\infty}^{\infty}d\omega B_{\text{cav}_{x}}(\omega)(t_{\text{out}})e^{-i\omega(t-t_{\text{out}})},\\
P_{u,\text{out}}(t) & =\frac{i}{\sqrt{2\pi}}\int_{-\infty}^{\infty}d\omega\beta_{u}(\omega)(t_{\text{out}})e^{-i\omega(t-t_{\text{out}})}
\end{align}
annihilate output fields produced as energy is dissipated from the
polaritonic system by bath modes at time $t_{\text{out}}>t$, and
the different $\beta_{u}$ are defined as $\beta_{u}(\omega)=\sum_{i=1}^{N_{u}}B_{u,i}(\omega)/\sqrt{N_{u}}$
for $u=\text{R},\text{RC}$ and $\beta_{u}(\omega)=\sum_{i=1}^{N_{\text{R}}}|\text{G}\rangle\langle B_{u,i}(\omega)|/\sqrt{N_{\text{R}}}$
for $u=cis,trans$. Notice that for $|\text{P}\rangle$, the Heisenberg-Langevin
equations (cf. equations (\ref{eq:pP}) and (\ref{eq:pPdone})), input-output
relations (cf. equations (\ref{eq:ioP}) and (\ref{eq:io0P})), and
the associated notation are identical to those in the bare case (Supplementary
Note \ref{subsec:reaction}). In contrast, the Heisenberg-Langevin
equation (\ref{eq:pRC1}) contains the third-order RC polarization
$P_{\text{RC},3}$ (defined in Methods), whose dynamics become relevant
when the polaritonic device is pump-excited.

We now treat the dynamics of $P_{\text{RC},3}$, beginning with evaluation
of the exact Heisenberg-Langevin equation for $a_{\text{RC},i}^{\dagger}a_{\text{RC},i}a_{\text{RC},i}$:\begin{subequations}\label{eq:3a}
\begin{align}
\frac{d[a_{\text{RC},i}^{\dagger}(t)a_{\text{RC},i}(t)a_{\text{RC},i}(t)]}{dt} & =\frac{da_{\text{RC},i}^{\dagger}(t)}{dt}a_{\text{RC},i}(t)a_{\text{RC},i}(t)+a_{\text{RC},i}^{\dagger}(t)\frac{da_{\text{RC},i}(t)}{dt}a_{\text{RC},i}(t)+a_{\text{RC},i}^{\dagger}(t)a_{\text{RC},i}(t)\frac{da_{\text{RC},i}(t)}{dt}\label{eq:chain}\\
 & =-i(\omega_{\text{RC}}-i3\gamma_{\text{RC}}/2+2\Delta)[a_{\text{RC},i}^{\dagger}(t)a_{\text{RC},i}(t)]a_{\text{RC},i}(t)-2ig_{\text{RC}}[a_{\text{RC},i}^{\dagger}(t)a_{\text{RC},i}(t)]c_{\text{RC}}(t)\label{eq:3aKEEP}\\
 & \quad+ig_{\text{RC}}[c_{\text{RC}}^{\dagger}(t)a_{\text{RC},i}(t)]a_{\text{RC},j}(t)-2i\Delta[a_{\text{RC},i}^{\dagger}(t)a_{\text{RC},i}^{\dagger}(t)a_{\text{RC},i}(t)a_{\text{RC},i}(t)]a_{\text{RC},i}(t),\label{eq:3aTHROW}
\end{align}
\end{subequations}where in going from equations (\ref{eq:chain})
to (\ref{eq:3aKEEP}) and (\ref{eq:3aTHROW}) we have used 
\begin{equation}
\frac{da_{\text{RC},i}(t)}{dt}=-i(\omega_{\text{RC}}-i\gamma_{\text{RC}}/2)a_{\text{RC},i}(t)-ig_{\text{RC}}c_{\text{RC}}(t)-2i\Delta a_{\text{RC},i}^{\dagger}(t)a_{\text{RC},i}(t)a_{\text{RC},i}(t),
\end{equation}
obtainable by inspection of equation (\ref{eq:pRC1}). We now analyze
the terms that contribute to the dynamics of the anharmonic-induced
polarization. As written above, all terms in equation (\ref{eq:3aKEEP})
and (\ref{eq:3aTHROW}) are proportional to the product of a quantity
in square brackets and a creation operator. In particular, the first
(second) term in equation (\ref{eq:3aKEEP}) is proportional to $[a_{\text{RC},i}^{\dagger}a_{\text{RC},i}]a_{\text{RC},i}$
($[a_{\text{RC},i}^{\dagger}a_{\text{RC},i}]c_{\text{RC}}$) and interpretable
as creation of a population in the first excited RC state followed
by an RC (RC cavity) transition. On the other hand, the first (second)
term in equation (\ref{eq:3aTHROW}) is proportional to $[c_{\text{RC}}^{\dagger}a_{\text{RC},i}]a_{\text{RC},i}$
($[a_{\text{RC},i}^{\dagger}a_{\text{RC},i}^{\dagger}a_{\text{RC},i}a_{\text{RC},i}]a_{\text{RC},i}$)
and interpretable as the creation of a RC-cavity coherence (population
of the second excited RC state) followed by an RC transition. Because
we supposed that the pump pulse only creates singly excited RC population
(in the form of dark RC states; see main text), we neglect both terms
in equation (\ref{eq:3aTHROW}) \cite{Ribeiro2018vp}. Moreover, we
approximate the $a_{\text{RC},i}^{\dagger}a_{\text{RC},i}$ in the
rightmost term of equation (\ref{eq:3aKEEP}) as $f_{\text{pump}}=\sum_{i=1}^{N_{\text{RC}}}a_{\text{RC},i}^{\dagger}a_{\text{RC},i}/N_{\text{RC}}$,
the effective fraction of total RC molecules populated via relaxation
from polariton to RC dark states during the delay time \cite{Ribeiro2018vp}.
With these steps, equation (\ref{eq:3a}) becomes equation (\ref{eq:pRC3})
in Methods. 

We associate absorbance into RC, absorbance into R, and reaction efficiency
$\eta$ with the fraction of the initial energy that is dissipated
by the nonradiative bath of $|\text{RC}\rangle$, nonradiative bath
of $|\text{R}\rangle$, and the bath inducing relaxation from $|\text{P}\rangle$
into the $trans$-HONO well, respectively. In analogy to using input-output
equations (\ref{eq:io0R}) and (\ref{eq:io0P}) to obtain equations
(\ref{eq:aR0b4}) for R absorption and (\ref{eq:eta0b4}) for reaction
efficiency $\eta_{0}$ in the bare case, we employ polaritonic input-output
equations (\ref{eq:ioz}) and (\ref{eq:ioP}) to write
\begin{align}
\text{absorbance}_{\text{RC}}(f_{\text{pump}},\omega) & =\gamma_{\text{R}}|S_{\text{RC}}(f_{\text{pump}},\omega)|^{2}\label{eq:absRC}\\
\text{absorbance}_{\text{R}}(f_{\text{pump}},\omega) & =\gamma_{\text{R}}|S_{\text{R}}(f_{\text{pump}},\omega)|^{2}\label{eq:absR}\\
\eta(f_{\text{pump}},\omega) & =\Gamma_{trans}|S_{\text{P}}(f_{\text{pump}},\omega)|^{2}.\label{eq:etaS}
\end{align}
The linear response functions $S_{\text{RC}}$, $S_{\text{R}}$, and
$S_{\text{P}}$ are defined such that each multiplied by $c_{\text{R},\text{in}}(\omega)$
yields $P_{\text{RC}}(\omega)$, $P_{\text{R}}(\omega)$, and $P_{\text{P}}(\omega)$,
respectively. From the Heisenberg-Langevin equations (\ref{eq:HL}),
it is evident that calculation of equations (\ref{eq:absRC})-(\ref{eq:etaS})
requires knowledge of $V_{\text{R}-\text{P}}$. As we do in Supplementary
Note \ref{subsec:reaction} for the bare case, we perturbatively treat
the IVR coupling in this Supplementary Note and neglect the terms
containing $V_{\text{R}-\text{P}}$ in equations (\ref{eq:absRC})
and (\ref{eq:absR}) for polaritonic spectra. The validity of this
treatment can be shown by noticing that
\begin{equation}
P_{\text{R}}(\omega)=\frac{g_{\text{R}}\sqrt{N_{\text{R}}}[\omega-\omega_{\text{P}}+i(\Gamma_{cis}+\Gamma_{trans})/2]}{(\omega-\omega_{\text{R}}+i\gamma_{\text{nonrad}}/2)[\omega-\omega_{\text{P}}+i(\Gamma_{cis}+\Gamma_{trans})/2]-V_{\text{R}-\text{P}}^{2}}c_{\text{R}}(\omega),
\end{equation}
recognizing that the fraction on the right-hand side is equal to $S_{\text{R},0}(\omega)$
(equation (\ref{eq:cR0b4})) up to a constant, and analyzing the roots
of the denominator of this fraction as done in the last paragraph
of Supplementary Note \ref{subsec:reaction}.

\begin{figure}[h]
\includegraphics[width=0.5\textwidth]{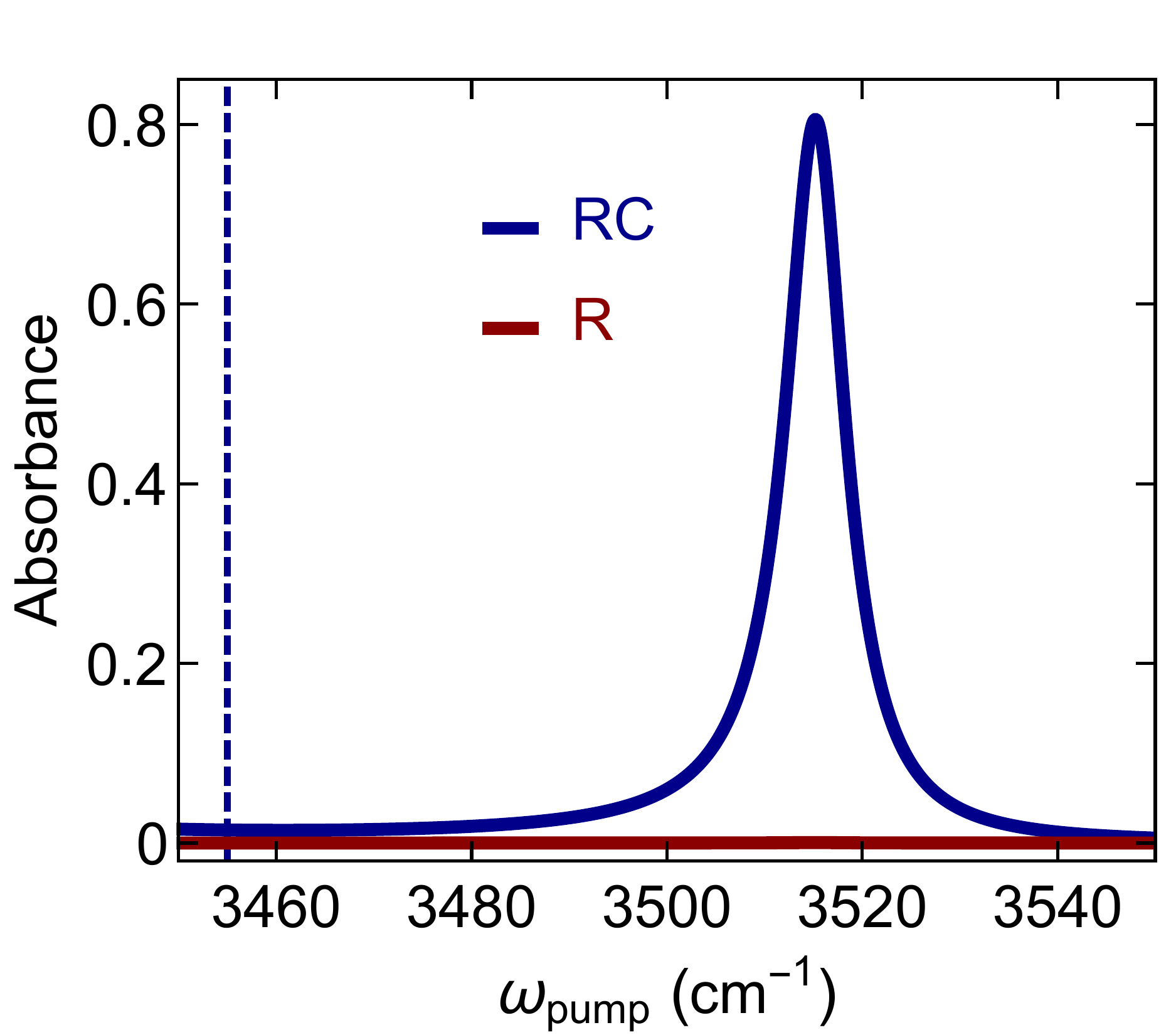}

\caption{\textbf{Linear absorption of pump pulse due to RC and R.} Spectra
of energy absorption into RC and R upon excitation of the highest
polariton (Supplementary Fig. \ref{fig:SI_rc_MF}b) with a pump pulse
impinging on the mirror of the RC cavity (Fig. \ref{fig:rc_scheme}b).
The dashed line indicates the energy of the bare OH stretch of RC
(in the absence of strong light-matter coupling). Accounting for the
lineshape broadening of the dark RC states and assuming this broadening
is equal to that ($\gamma_{\text{RC}}$, see Methods) of the bare
RC \cite{Du2018}, there should exist bath degrees of freedom that
mediate relaxation from the highest polariton to the dark RC states
in a matrix of Kr, whose Debye frequency is $\sim50\text{ cm}^{-1}$
\cite{KittelBook}.\label{fig:SI_rc_pump}}
\end{figure}

Calculation of linear absorption spectra upon pumping the RC can be
calculated in a fashion similar to equation (\ref{eq:io}). There
are two major procedural differences, which both arise in the Heisenberg-Langevin
equations (\ref{eq:HL}). For one, $f_{\text{pump}}$ is set to 0,
and thus equation (\ref{eq:pRC3}) for the third-order polarization
of RC can be neglected. Second, the only nonvanishing input operator
is that of the RC cavity, not the R cavity. This change corresponds
to dropping $c_{\text{R},\text{in}}$ in the Heisenberg-Langevin equation
(\ref{eq:cR}) for the R cavity and adding
\begin{equation}
c_{\text{RC},\text{in}}(t)=\frac{i}{\sqrt{2\pi}}\int_{-\infty}^{\infty}d\omega B_{\text{cav}_{\text{RC}}}(\omega)(t_{\text{in}})e^{-i\omega(t-t_{\text{in}})}
\end{equation}
to the right-hand side of the corresponding equation (\ref{eq:cRC})
for the RC cavity. Energy absorption of the pump pulse into R and
RC is still calculated via equations (\ref{eq:absR}) and (\ref{eq:absRC}),
respectively, and is plotted in Fig. (\ref{fig:SI_rc_pump}) for excitation
of the highest polariton. 

\subsection{Derivation of pump-dependent effective Hamiltonian for polaritonic
device\label{subsec:d_eff}}

In this Supplementary Note, we outline the derivation for the pump-dependent
effective Hamiltonian $H_{\text{system}}^{(\text{pump})}$ (equation
(\ref{eq:hpumphybrid})) for the polaritonic device. Although this
Hamiltonian was not used for calculations of absorption and reaction
efficiency, it provides an intuitive understanding of the associated
Heisenberg-Langevin equation (\ref{eq:HL}), on which these calculations
are based as described in Methods. Precisely, $H_{\text{system}}^{(\text{pump})}$
describes the projection of the Heisenberg-Langevin equation on the
system states, \textit{i.e.}, those acted on by $H_{\text{system}}$
(equation (\ref{eq:hS})), in the perturbative limit of $H_{\text{R}-\text{P}}$
(see equation (\ref{eq:hS})).

For reference, here we reproduce equation (\ref{eq:HL}) disregarding
$dP_{\text{P}}/dt$ and terms with $V_{\text{R}-\text{P}}$ and bath
operator $c_{\text{R},\text{in}}(t)$ (equation (\ref{eq:crin})):
\begin{align}
\frac{dP_{\text{RC}}(t)}{dt} & =-i(\omega_{\text{RC}}-i\gamma_{\text{RC}}/2)P_{\text{RC}}(t)\nonumber \\
 & \quad-ig_{\text{RC}}\sqrt{N_{\text{RC}}}c_{\text{RC}}(t)-2i\Delta P_{\text{RC},3}(t),\label{eq:pRC1e}\\
\frac{dP_{\text{RC},3}(t)}{dt} & =-i(\omega_{\text{RC}}+2\Delta-i3\gamma_{\text{RC}}/2)P_{\text{RC},3}(t)\nonumber \\
 & \quad-2ig_{\text{RC}}f_{\text{pump}}\sqrt{N_{\text{RC}}}c_{\text{RC}}(t),\label{eq:pRC3e}\\
\frac{dc_{\text{RC}}(t)}{dt} & =-i(\omega_{\text{cav}_{\text{RC}}}-i\kappa_{\text{RC}}/2)c_{\text{RC}}(t)\nonumber \\
 & \quad-ig_{\text{cav}}c_{\text{R}}(t)-ig_{\text{RC}}\sqrt{N_{\text{RC}}}P_{\text{RC}}(t),\label{eq:cRCe}\\
\frac{dc_{\text{R}}(t)}{dt} & =-i(\omega_{\text{cav}_{\text{R}}}-i\kappa_{\text{R}}/2)c_{\text{R}}(t)-ig_{\text{cav}}c_{\text{RC}}(t)\nonumber \\
 & \quad-ig_{\text{R}}\sqrt{N_{\text{R}}}P_{\text{R}}(t),\label{eq:cRe}\\
\frac{dP_{\text{R}}(t)}{dt} & =-i(\omega_{\text{R}}-i\gamma_{\text{R}}/2)P_{\text{R}}(t)\nonumber \\
 & \quad-ig_{\text{R}}\sqrt{N_{\text{R}}}c_{\text{R}}(t).\label{eq:pRe}
\end{align}
Define\begin{subequations}\label{eq:pRCp}

\begin{align}
P_{\text{RC}}' & =\frac{P_{\text{RC}}-P_{\text{RC},3}}{\sqrt{(1-2f_{\text{pump}})}}\\
 & =\frac{\sum_{i=1}^{N_{\text{RC}}}(1-a_{i}^{\dagger}a_{i})a_{i}}{\sqrt{(1-2f_{\text{pump}})N_{\text{RC}}}}\label{eq:pRCf}
\end{align}
\end{subequations}and\begin{subequations}\label{eq:pRCp3}
\begin{align}
P_{\text{RC},3}' & =\frac{P_{\text{RC},3}}{\sqrt{2f_{\text{pump}}}}\\
 & =\frac{\sum_{i=1}^{N_{\text{RC}}}(a_{i}^{\dagger}a_{i})a_{i}}{\sqrt{2f_{\text{pump}}N_{\text{RC}}}}.\label{eq:pRC3f}
\end{align}
\end{subequations}It is possible to check that 

\begin{equation}
i\partial_{t}\vec{v}=[H_{\text{system}}^{(\text{pump})}+H_{\text{system}}^{(\text{pump}),\text{non-Hermitian}}]\vec{v},\label{eq:generator}
\end{equation}
where $\vec{v}=(P_{\text{RC}}',P_{\text{RC},3}',c_{\text{RC}},c_{\text{R}},P_{\text{R}})$,
and $H_{\text{system}}^{(\text{pump})}$ and $H_{\text{system}}^{(\text{pump}),\text{non-Hermitian}}$
are respectively Hermitian and non-Hermitian parts,

\begin{align}
H_{\text{system}}^{(\text{pump})} & =\begin{pmatrix}\omega_{\text{RC}} & 0 & g_{\text{RC}}\sqrt{(1-2f_{\text{pump}})N_{\text{RC}}} & 0 & 0\\
0 & \omega_{\text{RC}}+2\Delta & g_{\text{RC}}\sqrt{2f_{\text{pump}}N_{\text{RC}}} & 0 & 0\\
g_{\text{RC}}\sqrt{(1-2f_{\text{pump}})N_{\text{RC}}} & g_{\text{RC}}\sqrt{2f_{\text{pump}}N_{\text{RC}}} & \omega_{\text{cav}_{\text{RC}}} & g_{\text{cav}} & 0\\
0 & 0 & g_{\text{cav}} & \omega_{\text{cav}_{\text{R}}} & g_{\text{R}}\sqrt{N_{\text{R}}}\\
0 & 0 & 0 & g_{\text{R}}\sqrt{N_{\text{R}}} & \omega_{\text{R}}
\end{pmatrix},\label{eq:hermitian}\\
H_{\text{system}}^{(\text{pump}),\text{non-Hermitian}} & =\begin{pmatrix}-i\gamma_{\text{RC}}/2 & i\gamma_{\text{RC}}\sqrt{2f_{\text{pump}}/(1-2f_{\text{pump}})} & 0 & 0 & 0\\
0 & -i3\gamma_{\text{RC}}/2 & 0 & 0 & 0\\
0 & 0 & -i\kappa_{\text{RC}}/2 & 0 & 0\\
0 & 0 & 0 & -i\kappa_{\text{R}}/2 & 0\\
0 & 0 & 0 & 0 & -i\gamma_{\text{R}}/2
\end{pmatrix}.\label{eq:anti-Hermitian}
\end{align}
Since only $H_{\text{system}}^{(\text{pump})}$ depends on parameters
of $H_{\text{system}}$, while only $H_{\text{system}}^{(\text{pump}),\text{non-Hermitian}}$
depends on dissipative parameters$\gamma_{x},\kappa_{x}$ for $x=\text{R},\text{RC}$,
equation (\ref{eq:generator}) provides the appealing effective physical
interpretation that $H_{\text{system}}^{(\text{pump})}$ governs the
coherent dynamics of the pumped system while $H_{\text{system}}^{(\text{pump}),\text{non-Hermitian}}$
characterizes its losses. In particular, we deduce \cite{Ribeiro2018vp}
that $P_{\text{RC}}'$ represents the 0 $\rightarrow$ 1 $|\text{RC}\rangle$
transition whose light-matter coupling reflects the combined amplitudes
of ground-state absorption and stimulated emission, whereas $P_{\text{RC},3}'$
represents the 1 $\rightarrow$ 2 $|\text{RC}\rangle$ transition
whose light-matter coupling reflects the amplitudes of excited-state
absorption. Although this interpretation was heuristically obtained
in previous work \cite{Ribeiro2018vp}, equations (\ref{eq:pRCp})
and (\ref{eq:pRCp3}) provide a formal justification for it. Additionally,
all other results in this work that follow from $H_{\text{system}}^{(\text{pump})}$\textemdash including
Figs. \ref{fig:rc_MF} and \ref{fig:SI_rc_MF} and associated discussion\textemdash are
for qualitative understanding only, precluding the need to ever specify
which of $P_{\text{RC}}$ or $P_{\text{RC}}'$ ($P_{\text{RC},3}$
or $P_{\text{RC},3}'$) these results pertain when referring to the
0 $\rightarrow$ 1 (1 $\rightarrow$ 2) $|\text{RC}\rangle$ transition.

\subsection{Representations of mixing fractions and energies for polaritonic
device\label{subsec:MF}}

In Supplementary Fig. \ref{fig:SI_rc_MF}, the plotted polariton mixing
fractions and energies are determined by diagonalization of system
Hamiltonian $H_{\text{system}}^{(\text{no pump})}$ (equation (\ref{eq:h0hybrid}))
for no pumping (a,b) or effective Hamiltonian $H_{\text{system}}^{(\text{pump})}$
(equation (\ref{eq:hpumphybrid}), see Supplementary Note \ref{subsec:d_eff}
for derivation and interpretation) for pumping (c). 

In Fig. \ref{fig:rc_MF}, color gradients and vertical positions of
the polariton line segments represent the mixing fractions and relative
energies, respectively, from Supplementary Fig. \ref{fig:SI_rc_MF}.
Specifically, {[}a, red and blue panels{]}, {[}a, purple panel, and
b, left panel{]}, and {[}b, right panel{]} of the former figure correspond
respectively to a, b, and c of the latter. To produce the (continuum)
color gradients in Fig. \ref{fig:rc_MF}, we use the following algorithms
to create significant resemblance to the discrete color gradients
in Fig. \ref{fig:SI_rc_MF}.

First consider the case without pumping. Denote the mixing fraction
of species $i=1,2,3,4$\textemdash corresponding to $|\text{cav}_{\text{RC}}\rangle,|\text{RC}\rangle,|\text{R}\rangle,|\text{cav}_{\text{R}}\rangle$,
respectively\textemdash for polariton $n=1,2,3,4$ as $f_{x}^{(i)}$,
where the polaritons are indexed from highest energy ($i=1$) to lowest
energy ($i=4$), and $\sum_{i}f_{i}^{(n)}=\sum_{n}f_{i}^{(n)}=1$.
Then we define quantities $g_{i}^{(n)}$ to indicate the positions
of gradient markers for polariton $i$ and species $x$:
\[
g_{i}^{(n)}=\begin{cases}
\text{round}\left(100\times(f_{i}^{(n)}+f_{i+1}^{(n)})/2\right) & i=1,2,3\\
\text{round}\left((100\times(f_{i}^{(n)}+1)/2)\right) & i=4
\end{cases}
\]
where $\text{round}(x)=\lceil x-0.5\rceil$ maps $x\in\mathbb{R}$
to the nearest integer (and odd multiples of 0.5 to the lower of the
two equally close nearest integers). For each $n$ and a gradient
scale that ranges from 0-100, a marker is at position $g_{i}^{(n)}\neq0$
with the color assigned to species $i$; if $g_{i}^{(n)}=0$, no marker
is placed. 

The case with pumping is similar. Due to the large anharmonicity of
$|\text{RC}\rangle$, the lowest-energy eigenstate of the effective
Hamiltonian equation (\ref{eq:hpumphybrid}) essentially represents
just the $|\text{RC}\rangle$ 1 $\rightarrow$ 2 transition. The remaining
eigenstates, which correspond to the polaritons, have insignificant
character of this transition (Fig. \ref{fig:SI_rc_MF}c). We therefore
combine the mixing fraction of the $|\text{RC}\rangle$ 1 $\rightarrow$
2 transition with that of $|\text{RC}\rangle$. The gradient markers
are then placed as described above for the case without pumping. 

\begin{figure}[th]
\includegraphics[width=0.5\textwidth]{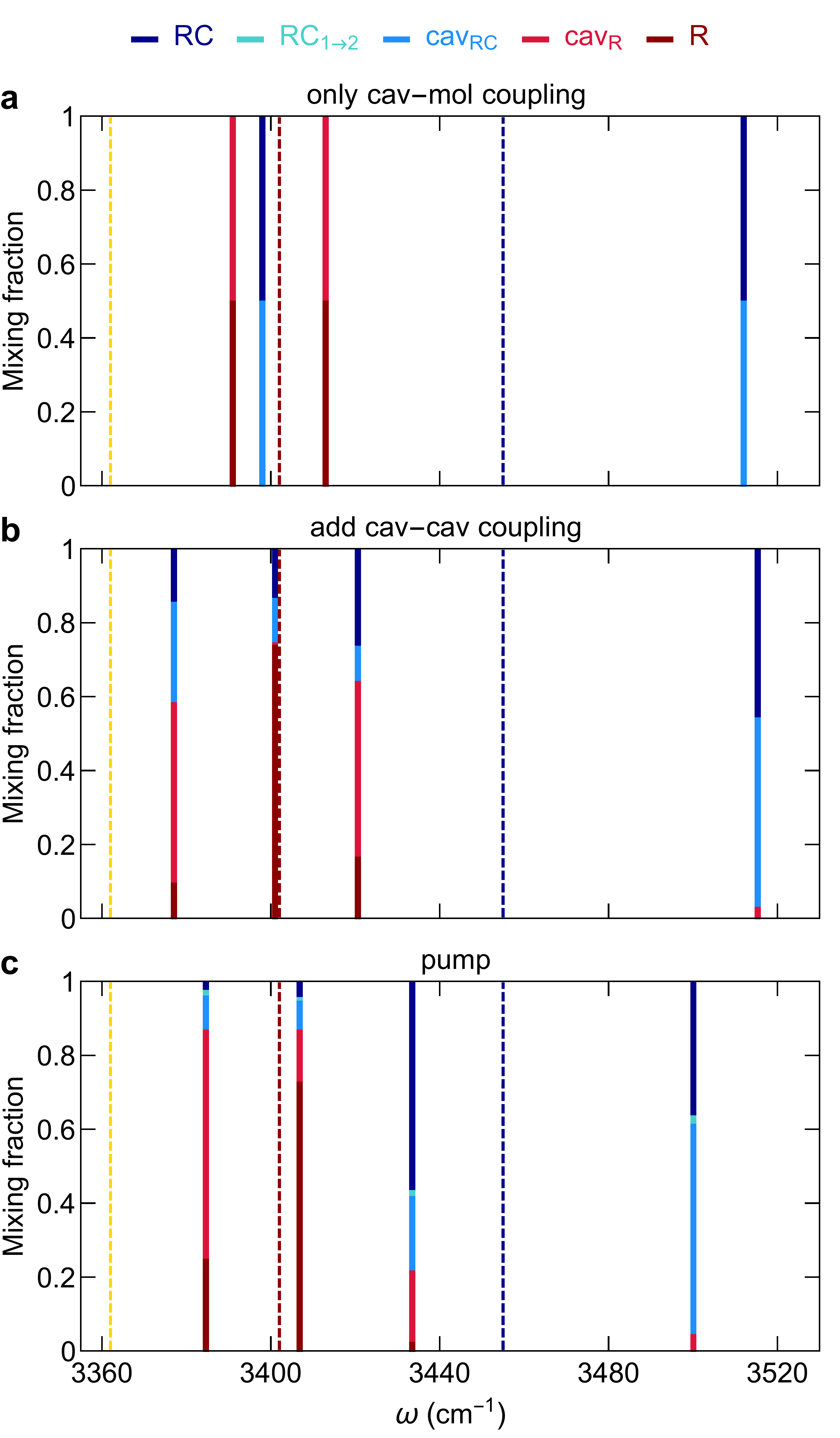}

\caption{\textbf{Effects of strong coupling and pumping on hybridization.}
The eigenstates and mixing fractions of the polaritonic device (Fig.
\ref{fig:rc_scheme}) under different circumstances for parameters
used in the simulations of Fig. \ref{fig:rc_ON}: \textbf{a}, with
strong coupling between each cavity and the 0 $\rightarrow$ 1 OH
stretches of its hosted molecules, \textbf{b}, with cavity-molecule
and cavity-cavity strong coupling, and \textbf{c}, after a pump pulse
acting on the system described by \textbf{b} excites 30\% of RC molecules
into the $v=1$ OH state. In each plot, the dashed lines indicate
the bare energies of the 0 $\rightarrow$ 1 vibrational transitions
of RC (dark blue) and reactant (R, dark red), as well as that of the
quantum state $|\text{P}\rangle$ (yellow) that first receives population
from $|\text{R}\rangle$ via IVR and then relaxes into P states (see
main text).\label{fig:SI_rc_MF}}
\end{figure}

\end{document}